\documentclass{aa}

\usepackage{graphicx}
\usepackage{txfonts}
\usepackage{longtable,booktabs}
\usepackage[breaklinks=true]{hyperref}
\usepackage{xcolor}
\hypersetup{
    colorlinks,
    linkcolor={blue!50!black},
    citecolor={blue!50!black},
    urlcolor={blue!50!black}
}
\usepackage{natbib}

\usepackage{ulem} 

\newcommand{\Msun}{\rm M_{\sun}}

\newcommand{\mdotbondi}{\dot{M}_{\rm Bondi}}
\newcommand{\mdotedd}{\dot{M}_{\rm Edd}}

\newcommand{\kpc}{\rm kpc}

\newcommand{\lsim}{\mathrel{\hbox{\rlap{\lower.55ex\hbox{$\sim$}} \kern-.3em\raise.4ex\hbox{$<$}}}}
\newcommand{\gsim}{\mathrel{\hbox{\rlap{\lower.55ex\hbox{$\sim$}} \kern-.3em\raise.4ex\hbox{$>$}}}}

\newcommand{\Ato}{A_{2} }

\newcommand{\Amax}{A_{\rm 2,max} }

\newcommand{\rbar}{r_{\rm bar} }

\newcommand{\hdisc}{h_{\rm disc, *}}

\newcommand{\tbar}{t_{\rm bar} }

\newcommand{\TNGF}{TNG50}

\defcitealias{rosasguevara2020}{RG20}
\defcitealias{rosasguevara2022}{RG22}

\begin{document}

   \title{Galaxy formation physics behind bar formation: A view from cosmological hydrodynamical simulations}
   \titlerunning{Galaxy formation physics behind bar formation}

   \author{Yetli Rosas-Guevara\inst{1}\thanks{email:yetli.rosas@dipc.org}
          \and
          Silvia Bonoli \inst{1}\fnmsep\inst{2}
          \and
          Ewald Puchwein\inst{3}
          \and
          Massimo Dotti \inst{4} \fnmsep\inst{5}\fnmsep\inst{6}
          \and
          Sergio Contreras \inst{1}
          }

  \institute{Donostia International Physics Centre (DIPC), Paseo Manuel de Lardizabal 4, 20018 Donostia-San Sebastian, Spain
   \and
             IKERBASQUE, Basque Foundation for Science, E-48013, Bilbao, Spain
    \and
            Leibniz-Institut fur Astrophysik Potsdam, An der Sternwarte 16, 14482 Potsdam, Germany
    \and
             Dipartimento di Fisica G. Occhialini, Universit\`{a} di Milano-Bicocca, Piazza della Scienza 3, IT-20126 Milano, Italy
     \and
             INFN, Sezione di Milano-Bicocca, Piazza della Scienza 3, IT-20126 Milano, Italy
    \and
            INAF, Osservatorio Astronomico di Brera, Via E. Bianchi 46, I-23807, Merate, Italy
            }

   \date{Received September 15, 1996; accepted March 16, 1997}


  \abstract
{We present a suite of zoom-in cosmological simulations of Milky Way-like galaxies with a prominent disc component and a strong bar in their centre, based on a subsample of barred galaxies from the TNG50 magneto-hydrodynamic simulation. We modify the physical models that regulate star formation, namely, supernova feedback and black hole quasar feedback, to examine how they affect the disc and bar formation. We find that, independently of the feedback prescriptions, all galaxies show a similar morphology, which is dominant in comparison with the bulge mass. The black hole quasar feedback models used in this study do not affect bar formation, although they can affect the bar strength and length. The energy released by the supernovae causes a delay in the time of bar formation and, in models with the strongest feedback, galaxies form stable discs against bar formation. This could be understood since supernova feedback influences disc and bulge assembly, resulting in discs with lower mass content, radial velocity dispersion and larger size as the supernova feedback strength increases. We study disc stability using three bar instability criteria proposed in the literature. We find that galaxies with varied supernovae and black hole quasar feedback satisfy these criteria at the moment of bar formation, except in extreme cases where the galaxy lacks or has weak supernova feedback. In these models, two of three criteria fail to forecast the existence (or absence) of a bar, probably because they do not account for the influence of a massive and compact bulge. Our findings provide insights into the physical processes behind bar formation and highlight the importance of additional conditions, other than a massive and compact disc that promote bar formation.}



   \maketitle
%

\section{Introduction}

 Galactic bars are elongated structures of stars, gas, and dust that extend from the centres of many disc galaxies, including the Milky Way. They are common features observed in spiral galaxies across the Universe; at least two-thirds of the local spirals have a bar (e.g. \citealt{sellwood1993,masters2011,gavazzi2015}). Bars are also observed in high redshift galaxies, although the fraction is still uncertain \citep[e.g.][]{guo2023,leconte2023,guo2024}.
These bars form due to gravitational instabilities within the galactic disc, which lead to the redistribution of angular momentum (e.g. \citealt{athanassoula2002,athanassoula2003,debattista2006,sellwood2012}). This process could funnel interstellar gas towards the galactic centre, often resulting in an enhanced central bulge and increased star formation activity in the core regions \citep[e.g.][]{spinoso2017, donohoe2019,george2019}. Stellar bars play an active role in the secular evolution of galaxies by redistributing mass and angular momentum, thereby influencing the overall morphology and dynamics of the galaxy.

One open question is which disc galaxies form a galactic bar and when. Many studies have investigated the conditions of a disc to form a bar since the early $60$s, with the seminal work of \cite{toomre1964}, which emphasises the importance of random stellar motions in maintaining the stability of galactic discs and provides a framework for understanding the dynamics of stellar systems. \cite{toomre1964} proposed a dimensionless parameter, the Toomre parameter $Q_{\rm T}$, that is employed to evaluate the stability of a differentially rotating disc:
\begin{equation}
Q_{T}= \frac{\kappa \sigma_{\rm R}}{3.36 G \Sigma},
\label{eq:Toomre}
\end{equation}
where $\kappa$ and $\sigma_{\rm R}$ denote the epicyclic frequency and velocity dispersion of the stars in the radial direction, respectively; $\Sigma$ denotes the face-on stellar surface density profile; and $G$ denotes the gravitational constant. The disc stability in response to axisymmetric density perturbations is determined by the Toomre parameter. If $Q_{\rm T}<1$, the disc is unstable.  \cite{ostriker1973} proposed that the ratio between the kinetic energy of rotation to the total gravitational energy of the galaxy has to be $\lsim 0.14$ to be stable. 
In addition, \citealt[][(ELN)]{efstathiou1982}, by studying models with exponential surface density profiles and flat rotation curves at large radii, investigated the formation of bars. The authors find that a critical criterion for stability in bar-like modes in disc galaxies is given by
 \begin{equation}
     \epsilon_{\rm ELN} \equiv v_{\rm Max}/(M_{\rm disc}G/r_{\rm disc})^{1/2}\geq 1.1,
\label{eq:ELN}
 \end{equation}
where $v_{\rm Max}$ is the maximum rotational velocity of the system, $r_{\rm disc}$ corresponds to the scale length of the exponential disc and $M_{\rm disc}$ is the total disc mass. They also explore the role of a hot halo component in increasing this ratio to stabilise the galaxy against bar formation. The study highlights the importance of balancing parameters like rotational velocity, scale length, and total disc mass to maintain stability in disc galaxies.  \cite{izquierdo2022} have investigated if the ELN criterion is satisfied for massive galaxies with strong bars in the TNG100 and TNG50 simulations. They find that more than 70 per cent of the barred galaxies satisfied the ELN criterion, while in galaxies that do not satisfy the criterion, it is because the bar formation is triggered by a tidal interaction rather than secular evolution.
This criterion was revised by \citealt[][(MMW)]{mo1998} with a focus on more fundamental properties, including the spin parameter ($\lambda$), the stellar disc mass fraction ($m_{\rm disc}=M_{\rm disc}/M_{h}$), and the angular momentum of the disc ($J_{\rm disc}$) and halo ($J_{\rm h}$). The authors found in their disc and halo profile that the discs are stable if $\lambda_{\rm MMW}>\lambda_{\rm crit}$ where $\lambda_{\rm MMW}$ is defined as
\begin{align}
   \lambda_{\rm MMW}&\equiv \lambda\frac{J_{\rm disc}}{J_{\rm h}}/\frac{M_{\rm disc}}{M_{h}} \nonumber\\
     &\gtrsim \lambda_{\rm crit} = \sqrt{2} \epsilon_{\rm ELN}^2 m_{\rm disc} f_{c}^{1/2} f_{\rm R}^{-1} f_{\rm V}^{-2}, \label{eq:MMW}
\end{align}
 where $\lambda=\mid\!\! \vec J_{h}\!\!\mid \!\!/( \sqrt{2} M_{200}V_{200}r_{200})$ where  $\vec J_{h}$ is the angular momentum  evaluated at $r_{200}$ \citep{bullock2001}. The functions $f_{c}, f_{\rm R}, f_{\rm V}$ weakly depend on the profile models of the disc and halo as seen in Fig. 3 by \citealt[][(MMW)]{mo1998}. Because of this, a useful but rough approximation is that $\lambda_{\rm MMW}>m_{\rm disc}$ that could be translated into $\lambda\frac{j_{\rm disc}}{j_{\rm h}}/\frac{M_{\rm disc}}{M_{h}}>1$ where $j_{\rm disc}$ and $j_{\rm h}$ are the specific angular momentum of the disc and halo.

Although these criteria are widely used to predict bar instabilities in galaxy formation models,  some theoretical and observational studies have shown that bar formation could be more complex and that other ingredients may affect bar evolution. For instance, \cite{athanassoula2008} discusses that ELN does not take into account the (stabilising) effect of the disc velocity dispersion or
the central concentration. \cite{ghosh2023} found that using different contributions of the thick discs, some of their galaxies do not satisfy the ELN criterion. \cite{romeo2023} found that for more diverse barred/unbarred disc galaxy samples (the galaxies of types S0–BCD  from the Spitzer Photometry and Accurate Rotation Curves' sample (SPARC) and galaxies of type Im from the Local Irregulars That Trace Luminosity Extremes The HI Nearby Galaxy Survey (LITTLE THINGS), some criteria could not be applied for all the diversity of galaxies found in the local Universe.
In addition, other components, such as the bulge, could avoid the formation of a bar. \cite{saha2018} show that disc galaxies with a bulge denser than their disc can prevent bar formation even though the disc is maximal or unstable to bar instabilities. This is due to the rapid loss of angular momentum and rapid heating in the centre of initially strong bar and spiral arms. Also, high-rotating dark matter haloes could prevent the growth of bar instabilities. \cite*{saha2013} also show that bar formation is favoured in haloes corotating with the disc with a dark matter spin parameter ($\lambda_{\rm DM}$) in the range of  0 and 0.07.
However, \cite{long2014} have shown that the bar growth is halted in haloes with larger spins ($\lambda_{\rm DM}\gsim0.03$). \cite{collier2018} suggested that not only the spin parameter but also the shape of the dark matter halo is important to the evolution of bars. \cite{yurin2015} have considered the stability of discs by inserting already formed stellar discs in haloes using the Aquarius simulation project that consists of zoom-in, dark matter-only simulations. These authors found that the 3-D shape of the dark matter halo could affect the stability of the disc. Independently of the results, all these studies show the importance of the angular momentum exchange between the dark matter and the stellar components in the formation of a bar (e.g. \citealt{sheth2012,athanassoula2013}).

The bulge and disc structural properties are related to the formation of galactic bars and are the result of the formation history of the galaxy in the first place \citep{mo1998,rosito2019,bi2022a,joshi2024}. One crucial driver that could influence the morphology of galaxies since early times is star formation feedback and AGN (Active Galactic Nuclei) feedback. Supernova (SN) feedback refers to the impact of supernova explosions on the interstellar medium (ISM). Massive stars end their lives in supernova explosions, releasing vast amounts of energy, radiation, and heavy elements. This feedback process has several critical effects on the ISM: it injects energy, creating shock waves that can compress or disperse gas clouds; it enriches the ISM with heavy elements necessary for new star formation; and it regulates the star formation rate by preventing runaway star formation that could deplete the galaxy gas reservoir. AGN feedback is powered by the accretion of gas into the supermassive black holes that are thought to be located in the centres of all (sufficiently large) galaxies and has been proposed to mostly affect star formation in massive galaxies \citep[e.g.][]{springel2005,bower2006,croton2006}.
Both SN and AGN feedback are critical processes in all the galaxy formation models today \citep[e.g.][]{springel2005,bower2006,croton2006}. Both processes inject energy and momentum into the ISM, potentially influencing the star formation and regulating the galaxy build-up and, subsequently, the bar formation. In particular,  SN feedback has been demonstrated to be efficient in the accumulation of central low angular momentum gas, allowing for the build-up of cold discs with flat rotation curves comparable with observations \citep[e.g.][]{navarro1991}.

Barred galaxies have started to be explored in a fully cosmological context recently, thanks to high-resolution zoom-in simulations (e.g. \citealt{kraljic2012,scannapieco2012,bonoli2016}). In particular, \cite*{bonoli2016} present a zoom-in simulation of a Milky Way-type galaxy,  \textit{ErisBH}, a sibling of the \textit{Eris} simulation \citep{guedes2011}, but with black hole (BH) subgrid physics included. \cite{bonoli2016} and \cite{spinoso2017} find that the simulated galaxy forms a strong bar below $z\sim1$, and the authors point out that the disc in the simulation is more prone to instabilities compared to the original \textit{Eris}, possibly because of early AGN feedback affecting the central part of the galaxy. \cite{zana2019}, studying a larger suite of \textit{Eris}, highlights the effects of the feedback processes on the formation time and final properties of the bar. Recently, \cite{fragkoudi2020,fragkoudi2024} have studied barred galaxy formation in the Auriga suite, which consists of high-resolution, magnetohydrodynamical cosmological zoom-in simulations of galaxy formation \citep{grand2017}. The authors find that the simulated galaxies that can reproduce many chemodynamical properties of the stellar populations seen in the Milky-way bulge have quiet merger histories. Auriga simulations have also been employed by \cite{irodotou2022} to determine that AGN feedback can influence the final properties of a bar, but not its formation.  The formation of bars in high-redshift counterpart galaxies of the local spiral galaxies formed in dense environments and with two distinct SN feedback subgrid models has been investigated by \cite{bi2022b} using high-resolution zoom-in simulations. The authors find that the properties and evolution of galactic bars are significantly impacted by the varying SN feedback. These changes are associated with the interactions and cold accretion processes, which were influenced by the type of feedback implemented.

With cosmological hydrodynamic simulations (\citealt{vogelsberger2014a,schaye2015,pillepich2018b,nelson2018}, see also the recent review of \citealt{vogelsberger2020a, crain2023}), it has been possible to follow the formation and evolution of the barred galaxy population in a cosmological context and with a significant statistical sample. Analysing the EAGLE simulation, \cite{algorry2017} find that bars slow down quickly as they evolve, expanding the inner parts of the dark matter halo. \cite{rosasguevara2020,rosasguevara2022} study massive barred disc galaxies at $z=0$ in the TNG100 and TNG50 simulation (see also \citealt{peschken2019,zhao2020,zhou2020} for Illustris and IllustrisTNG), finding that barred galaxies are less star-forming and more gas poor than unbarred galaxies. Following the evolution of barred galaxies back in time, \cite{rosasguevara2020} find that these objects assembled most of their disc components before bar formation and at earlier times than unbarred galaxies (see also \citealt{izquierdo2022}).

Even though these simulations have shown the possibility of understanding the evolution and formation of bars in a cosmological context, there are many complex galaxy formation processes that are taking place, and it is difficult to disentangle all the relevant processes. Taking advantage of previous works, we perform zoom-in simulations of six Milky Way-like galaxies from the TNG50 simulations. These galaxies have a stable galactic bar at $z=0$ and formed between $z=3$ and $z=1.5$ (more than $8$  billion years ago) and with quiet merger histories and isolated haloes. These zoom-in simulations are run with the \textsc{Arepo} code \citep{springel2010} and similar initial conditions as the original TNG50 simulation \citep{pillepich2019,nelson2019b} in the sense that the large-scale tidal field is very similar, but only the galaxy region simulated with the same resolution as TNG50.
Our primary goal is to study the interplay of galaxy formation physics with the formation and evolution of galactic bars. For that, we use the same galaxy formation physics except for variations in the supernova and AGN feedback models to try to identify the specific impacts of these feedback channels. We also explore how these variations impact different bar instability criteria used in the field: the \citealt{toomre1964}, and  \citealt[][(ELN)]{efstathiou1982}, and \citealt[][(MMW)]{mo1998} criteria.

The paper is structured as follows. In section~\ref{sec:method}, we introduce our selection of the disc galaxies and the methodology for identifying a bar and dynamics components, as well as give a brief overview of the supernova and AGN feedback subgrid physics of the TNG50 and their variations used here. In section~\ref{sec:galprop}, we study the redshift evolution of their bars and their host galaxy components. We study the criteria of bar instability and how they are affected by the different galaxy formation models in section~\ref{sec:barformation}. Finally, in sections \ref{sec:discussion} \& \ref{sec:summary},  we discuss and summarise our findings, respectively.

\section{Methodology}
\label{sec:method}
In this section, we describe the object selection criteria, the generation of the initial conditions, and the TNG galaxy formation model. We especially describe the variations of the galaxy formation model. This project accounts for $42$ resimulations, with six haloes, each with seven different variations of the galaxy formation model. These simulations were performed with the moving-mesh \textsc{Arepo} code \citep{springel2010}, combining Tree-PM and Godunov/finite volume methods to discretise space. The quasi-Langragian scheme is second-order in space and time.

The cosmology parameters adopted are from the \cite{planck2016} Cosmology:
$\Omega_\Lambda=0.6911$, $\Omega_{\rm m}=0.3089$, $\Omega_{\rm b}=0.0486$, $\sigma_8=0.8159$, $h=0.6774$, and $n_{s}=0.9667$  where  $\Omega_\Lambda$, $\Omega_{\rm m}$, and  $\Omega_{\rm b}$ are the average densities of dark energy, dark matter and baryonic matter in units of the critical density at $z=0$,  $\sigma_8$ is the square root of the linear variance,  $h$ is the Hubble parameter ($H_{0}\equiv h \,100\, \rm km \, s^{-1}$) and, $n_{s}$  is the scalar power-law index of the power spectrum of primordial adiabatic perturbations.

The suite of zoom-in simulations has a similar resolution to the \TNGF~ simulation, whose particle mass resolution is $4.5\times 10^5 \Msun$ for dark matter particles, whereas the mean cell mass resolution is $8.5\times10^4\Msun$ for gas. A comparable initial mass is passed down to stellar particles, which subsequently lose mass through stellar evolution.
The spatial resolution, i.e. the gravitational softening, for collisionless particles (dark matter and stellar particles) is $575$ comoving pc down to $z=1$, after which it remains constant at $288$ pc in physical units down to $z=0$. In the case of the gas component, the gravitational softening is adaptive and based on the effective cell radius, down to a minimum value of $72$ pc in physical units, which is imposed at all times.

Galaxies and their haloes are identified as bound substructures using a \textsc{fof} (Friends-of-Friends) and then the \textsc{subfind} algorithm \citep{springel2001} and tracked over time by the \textsc{Sublink} merger tree algorithm \citep{rodriguezgomez2015}.
Halo masses ($M_{200}$) are defined as all matter within the radius $R_{200}$ for which
the enclosed mean density is $200$ times the critical density of the Universe.
In each \textsc{fof} halo, the central galaxy (subhalo) is the first (most massive) subhalo of each \textsc{fof} group. Its satellites are the remaining galaxies within the \textsc{fof} halo. The stellar mass of a galaxy is defined as all the stellar matter assigned to host subhaloes.

\subsection{Object selection}

\begin{figure*}
\includegraphics[width=2.05\columnwidth]{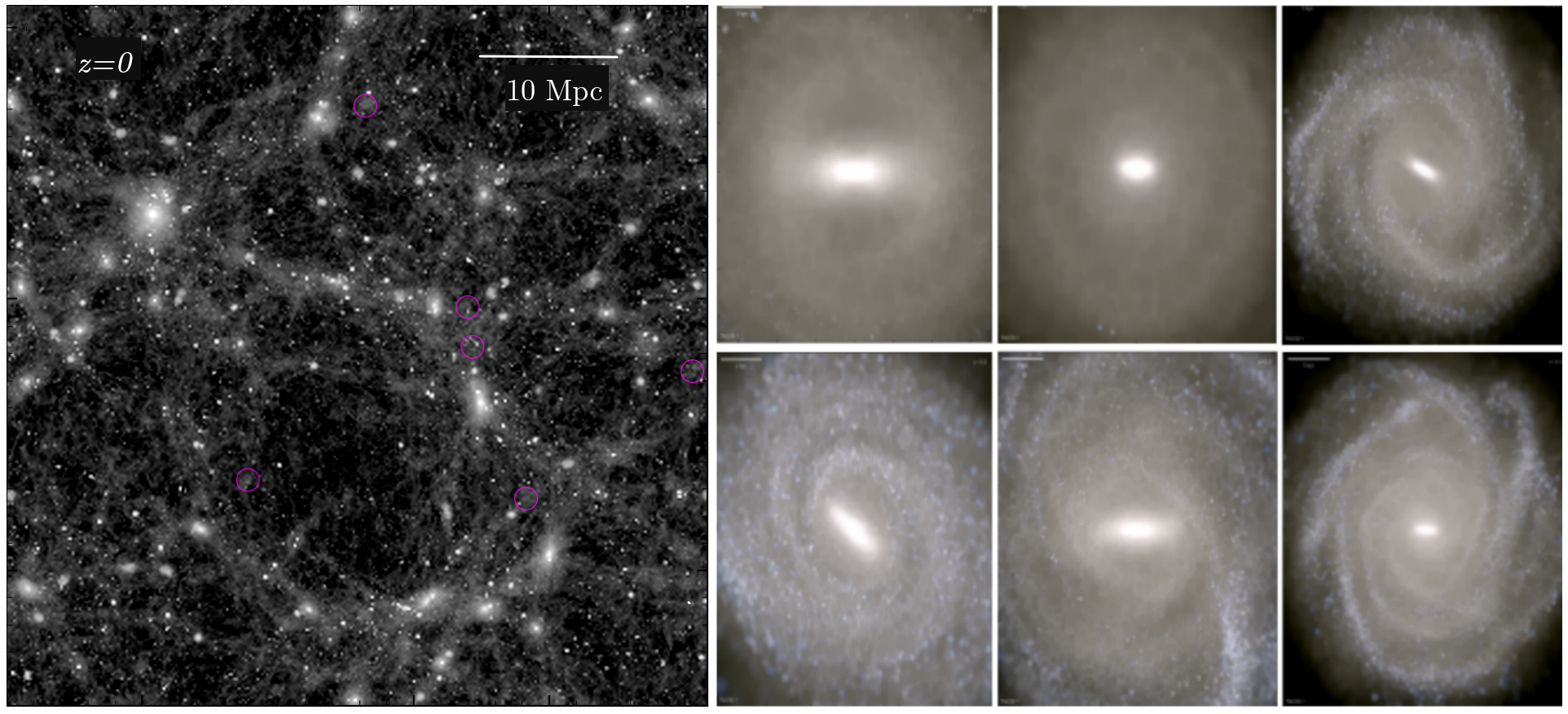}
\caption{Left panel: Dark matter distribution of the TNG50 simulation. Magenta empty circles show the position of the selected haloes of the zoom-in simulations. Right panel: Mocked images in JWST  NIRCam F200W, F115W, and F070W filters (face-on) calculated in \protect\citealt[][(TNG database)]{nelson2018} for the TNG50 galaxies that are resimulated. The NIRCam blue channel highlights the young population of the galaxy, and the NIRCam red channel older populations. The original TNG50 galaxies present a massive disc with a smaller bulge and form a strong bar.}
\label{fig:DensityMap}
\end{figure*}

The suite of our zoom-in simulations is based on barred galaxies formed in the \TNGF~ simulation. For this, we use the bar catalogue of \cite*[][herein RG22]{rosasguevara2022} and select central disc galaxies with a bar at $z=0$. We select those whose dark matter haloes with masses  ($M_{\rm 200}$) between $10^{11.5}$ and $10^{12} \Msun$  comparable to the Milky-Way like halo, yielding $31$ galaxies.
Furthermore, we concentrated on relatively isolated haloes. In particular, we used the tidal parameter that is defined as $\tau = M_{\rm 200, i}/M_{\rm 200, tar} (R_{\rm 200,tar}/R_{i})^3$ \citep{dahari1984}, where $M_{\rm 200,i}$ and  $M_{\rm 200, tar}$ are the halo masses of the closest massive neighbour and the target halo, respectively. $R_{200}$ is the size of the target haloes and $R_{i}$ is the distance to the closest massive neighbour.
We calculate $\tau$ of our possible haloes candidates in order to assess the interaction strength between our candidates and their closest massive companion. This allows us to choose MW-like haloes that are isolated in the TNG50 dark matter simulation.  We calculated the tidal parameter for our halo candidates and selected those with values lower than $10^{-9}$. This condition corresponds to those haloes whose closest massive neighbour galaxy is smaller than the candidate halo ($M_{\rm 200,tar}$) and at a distance of $10 R_{\rm 200,tar}$ (see \citealt{grand2017}). We also verify that these galaxies have relatively quiet merger histories (no more than two major mergers in $z<2$). We select six dark matter halos from the \TNGF~ simulations \citep{nelson2019a} whose baryonic counterpart is a disc galaxy with a stellar mass $>10^{10}\Msun$ and have a small bulge ($B/T$ lie between $0.12$ and $0.25$). These galaxies hold a stable bar at $z=0$ that has formed between $z=1.5$ and $z=3$. The bar sizes lie between $2$ and $7$ kpc, and the disc sizes between $2.5$ and $6$ kpc. Most of the bars are strong bars ($\Amax>0.3$, see Appendix \ref{app:decomp} for bar identification method). The right panel of  Fig.~\ref{fig:DensityMap} shows mock images \citep[taken from TNG database][]{pillepich2019} in JWST NIRCam  F200W, F115W, and F070W filters of the six barred galaxies from the TNG50 that were resimulated in this work and Table~\ref{table:haloes} shows some properties of the bars and discs at $z=0$. We note that $5$ of the $6$ haloes belong to the catalogue of Milky Way analogues from \TNGF~ \citep{pillepich2023}. The left panel of Fig.~\ref{fig:DensityMap} depicts the dark matter density map of the \TNGF~ simulation, and the magenta circles mark the locations of the selected haloes which are not located in the densest regions of the cosmic web.
\begingroup
\renewcommand{\arraystretch}{1.5}
\begin{table*}
\caption{TNG50 galaxy sample at $z=0$. From left to right: Galaxy ID from TNG50 simulation,
the logarithm of halo mass, disc (thin and thick components) to total mass fraction from kinematic decomposition (from fitting surface density stellar profiles as described in Appendix  \ref{app:decomp}), bulge (classical component) to total mass fraction from kinematic decomposition (from fitting surface density stellar profiles), the effective radius from the bulge calculated from a Sersic profile, the disc scale length (fitting face-on stellar surface density profiles of the kinematic component), bar strength and bar length (using Fourier decomposing face-on stellar surface density as described in Appendix \ref{app:bars}.}

\label{table:haloes}
\centering
\begin{tabular}{cccccccccc} 
\hline
Galaxy ID  &  ${\rm log_{10}}(M_{\rm crit200})$     &   ${\rm log_{10}}(M_{*})$  & $D/T$ ($(D/T)_{\rm pso}$)  & $B/T$ ($(B/T)_{\rm pso}$) & $R_{\rm eff, bulge}$ &  $h_{\rm disc}$ & $A_{2,\rm max}$ & $r_{\rm bar}$ \\

           &   log$_{10}(M_{\odot})$                           &  log$_{10}(M_{\odot})$   & & & $\kpc$  &  $\kpc$ & & $\kpc$&\\
\hline
\hline
560751    &  $11.90$ &  $10.62$     &    $0.53$ 	 ($0.69$)  &	$0.25$ 	 ($0.29$)  &  $0.92$  &	 $4.52$    &	 $0.57$ 	& $6.48$
  \\
563732    &  $11.91$ &  $10.56$     &    $0.58$ 	 ($0.75$) &	 $0.26$ 	 ($0.12$)  & $0.81$  & 	 $5.51$ &	 $0.55$ &	 $2.64$  \\  
569251    &  $11.84$ &  $10.60$     &    $0.85$ 	 ($0.84$) &	 $0.07$ 	 ($0.11$)  &  $0.91$ &	$3.58$ &	 $0.61$ & 	 $3.52$ \\   
574286    &  $11.71$ &  $10.34$     &    $0.68$ 	 ($0.74$) &	 $0.11$ 	 ($0.20$)  &  $0.42$ &	$2.56$ &	 $0.53$ &	 $2.40$
  \\   
547293    &  $11.96$ &  $10.73$    &    $0.81$ 	 ($0.71$) &	 $0.05$ 	 ($0.24$)  &  $1.10$ &	 	 $2.72$  &	 $0.48$ & 	 $5.28$
 \\
543376    &  $11.97$ &  $10.79$     &   $0.67$ 	 ($0.85$) &	 $0.12$ 	 ($0.13$)  &  $0.90$ &	  	 $5.78$  &	 $0.38$ &	 $2.16$ \\
\bf{Median}    &  $\bf{11.91}$ &  $\bf{10.61}$     &   $\bf{0.67}$ ($\bf{0.73}$) &	 $\bf{0.12}$ ($\bf{0.17}$)  &  $\bf{0.91}$ &	  	 $\bf{4.05}$  &	 $\bf{0.54}$ &	 $\bf{3.08}$ \\
\hline
\end{tabular}
\end{table*}
\endgroup
\subsection{Initial Conditions}
The initial conditions for the zoom-in simulation suite were created by selecting all dark matter particles in the \TNGF~ parent simulation at $z=0$ that are located within $3 \times R_{200}$ (or $5 \times R_{200}$ in cases where the former was not sufficient to get contamination free objects) of the centres of the chosen objects. The particles were then traced back to the starting redshift of the \TNGF~ simulation to obtain the Lagrangian region of each object. Next, a tree structure was used to produce the unperturbed particle distribution for each zoom-in simulation. Within the Lagrangian region, the tree was refined down to the same resolution as the initial particle distribution in the \TNGF~ simulation. Then layers of lower resolution boundary particles were added, with the resolution set based on the distance to the nearest edge of the high-resolution region, until a base resolution (corresponding to 135 particles on a side) was reached that covers the remaining part of the \TNGF~ volume. Perturbations were added at $z=127$, the starting redshift adopted in the zoom-in simulations, using the Zeldovich approximation. Mode amplitudes and phases were drawn in Fourier space using the same random number seed and algorithm as in the \textsc{n-genic} code \citep{2015ascl.soft02003S}, so that perturbations consistent with the \TNGF~ parent simulation were imprinted. For lower resolution boundary particles, only modes up to the Nyquist frequency that corresponds to that resolution level were included. An updated version of the employed zoom-in initial conditions code will be presented in Puchwein et al. (in prep.).

Upon starting the \textsc{Arepo} simulations, the particles in the initial conditions file are split into dark matter and gas components and then diagonally displaced with respect to each other by half a diagonal particle spacing, while keeping the centre of mass unchanged. A uniform magnetic seed field with a comoving field strength of $10^{-14}$ Gauss is assumed at the initial redshift.

\subsection{The galaxy formation model and variations}
We used the IllustrisTNG galaxy formation model (The Next Generation) project\footnote{\cite{nelson2019a}; http://www.tng-project.org}  (\citealt{nelson2018,naiman2018,pillepich2018b,marinacci2018,springel2018}) and its predecessor  \textsc{Illustris} \citep{vogelsberger2013,vogelsberger2014a,vogelsberger2014b,genel2014,nelson2015,sijacki2015} with changes to star formation feedback, supermassive black hole growth, AGN feedback, and stellar evolution and chemical enrichment.

Gas radiative mechanisms are implemented with primordial \citep{katz1996} and metal line cooling, and heating by a time-dependent ultraviolet background field from stars and luminous AGN \citep{faucher2009}. Star formation in the dense interstellar medium is treated stochastically following an empirical Kennicutt-Schmidt relation \citep{springel2003}. Each stellar particle represents a population of stars with a common birth time following a Chabrier initial mass function.  The stellar evolution is modelled in order to calculate chemical enrichment and mass expelled into the interstellar medium due to AGB stars, SNIa and SNII. Also, the evolution and production of ten elements (H, He, C, N, O, Ne, Mg, Si, Fe, \& Eu) are individually tracked.

Previous studies have shown that physical processes that regulate star formation in galaxies can affect the formation of a bar (e.g. \citealt{bonoli2016,zana2019,irodotou2022}). Taking advantage of this, combined with the parametric study done for the TNG collaboration  \citep{pillepich2019}, we examine variations in physical processes that regulate the star formation of galaxies.
We can divide these variations into two categories: (1) those models that present modifications in the galactic winds generated from star formation, and (2) those models that present modifications in the BH physics of the model. The models of winds generated from star formation and AGN feedback will be explained in detail in the following subsections.

\subsubsection{The galactic wind model and its variations}
 For our suite of zoom-in simulations, we use the TNG model for galactic-scale, star formation-driven kinetic feedback \citep{pillepich2018b}. In this model, wind particles absorb the thermal energy of the inherited gas cell and are injected isotropically. The initial orientation of the wind particles is random and are launched with an initial speed that scales with the local, one-dimensional dark matter velocity dispersion $\sigma_{\rm DM}$ (as in Eq. (14) of \citealt{oppenheimer2006,oppenheimer2008,vogelsberger2013}), calculated with a weighted kernel over the $N=64$ nearest DM particles. In addition, the wind velocity has a dependency on redshift and a minimum value $v_{w, \rm min}$ given by
 \begin{equation}
     v_{w}={\rm max}[k_w \sigma_{\rm DM}(H_{0}/H(z))^{1/3},v_{w,\rm min} ],
 \end{equation}
 where $k_{w}$ is a dimensionless parameter. We note that the wind velocity depends on redshift, this ensures that wind velocity and virial halo mass increase in scale with redshift in the same way (see \citealt{pillepich2018b}). This choice of redshift-independent wind velocities at constant halo mass is motivated by semi-analytical model findings, where a comparable approach was required to recreate observed stellar mass functions and rest frame B- and K-band brightness functions throughout redshift \citep{henriques2013}.

 Once the wind injection velocity is known, the wind mass loading factor is determined by the specific energy available, $e_{w}$, which is connected to the energy released by SNII per produced stellar mass in the model. The TNG model includes two categories that affect the available wind energy: (i) some fraction of this energy is thermal, as determined by a parameter $\tau_{w}$; and (ii) wind energy is dependent on the metallicity of the star-forming gas cells, such that galactic winds are weaker in higher metallicity environments. The mass loading factor upon injection is calculated as follows:
\begin{equation}
    \eta_{w} = \dot{M}_{w}/\dot{M}_{\rm SFR}=\frac{2}{v^2_{w}}e_{w} (1-\tau_{w}),
\end{equation}
where $\dot{M}_{w}$ corresponds to the rate of gas mass to be converted into wind particles and $\dot{M}_{\rm SFR}$ the instantaneous, local, star formation rate.
Then the wind energy available from the star-forming gas cells depends on  metallicity $Z$ as follows:
\begin{equation}
    e_{w}= \overline{e_{w}} [f_{w,Z}+ \frac{1-f_{w,Z}}{1+Z/(Z_{w,\rm ref})^{\gamma_w,Z}}] N_{\rm SNII} E_{\rm SNII,51} 10^{51} \rm erg M_{\odot}^{-1},
\label{eq:energy}
\end{equation}
where $\overline{e_{w}}$ is a dimensionless free parameter of the
model, $E_{\rm SNII,51}$ denotes the available energy per core-collapse supernovae in units of $10^{51}$ erg. $N_{\rm SNII}$ is the number of SNII per formed stellar mass (in solar mass units) and depends on the shape of the IMF and the assumed minimum mass of core-collapse supernovae.  $f_{w,Z}$ is a parameter that reduces the energy at injection for gas cells with metallicities much larger than a given value $Z_{w,\rm ref}$.

We focus on the energy available from star-forming cells of the galactic winds over time in our suite of zoom-in simulations, which limits the intensity of the galactic wind. The energy is regulated by the dimensionless free parameter, $\overline{e_{w}}$, which is between $\overline{e_{w}}=0$ and $\overline{e_{w}}=7.2$. These values correspond to the No Wind (\textbf{NW}) model and the Strong Wind (\textbf{SW}) model, respectively. The value $\overline{e_{w}}=3.6$ is employed in the TNG50 simulation, and we refer to it here as the TNG50-like model (\textbf{TNG50-like}). In total, we have five modifications, including Weaker Winds (\textbf{WW}) and Medium Winds (\textbf{MW}). These variations are summarised in Table \ref{table:models}.

We remark that some of these variations have been explored to see their impact on the complete galaxy population in \cite{pillepich2018b}. We can see in their Fig. B1 how the global star formation history of a galaxy population can be affected by the strength of the winds.  It is also worth mentioning that we have tried other variations at low resolution, such as a higher fraction of the thermal energy in the wind ($\tau_{w}$) or higher wind velocity ($\kappa_{w}$), and we found that the effects on the star formation of a galaxy were similar as varying $\overline{e_{w}}$. However, at high redshift, the variations in the strength of the winds $\overline{e_{w}}$ affect the star formation density of the galaxy population (see Fig B1 of \citealt{pillepich2018b}).

\begingroup
\renewcommand{\arraystretch}{1.5}
\begin{table}
\caption{Variations in the \textsc{tng} galaxy formation model. Modified parameters in the galactic wind and quasar BH physics models. From left to right: Model name, wind strength parameter (Eq.\ref{eq:energy}), and BH feedback efficiency in quasar mode (Eq. \ref{eq:ehigh}). In the case of No wind/ No black hole models, all the parameters related to the model were set to 0.}
\label{table:models}
\centering
\begin{tabular}{cccccc}
\hline
Name            & $\overline{e_{w}}$ & $\epsilon_{f,\rm high}$   \\
               &                      &                         \\

\hline
\hline
NW (no winds)      &  $\textbf{--}$    &      $0.1$         \\
WW (weaker winds)   &  $\textbf{1.8}$     &   $0.1$        \\
TNG50-like          &  $\textbf{3.6}$     &   $0.1$     \\
MW (medium winds)     &  $\textbf{5.4}$     &  $0.1$      \\

SW (strong winds)   &  $\textbf{7.2}$     &  $0.1$         \\
NBH (no black holes)         &  $3.6$              &       $\textbf{--}$   \\
BHlowEff       & $3.6$               &   $\textbf{0.05}$   \\ 
\hline
\end{tabular}
\end{table}
\endgroup


\subsubsection{The BH model and its variations}
The other model used in our suite of zoom-in simulations is the subgrid physics of supermassive black holes (BHs) that was presented in \cite{weinberger2017}. The BHs are formed in massive haloes ($M>7.38 \times 10^{10}\Msun$) with an initial black hole mass of $1.18\times 10^{6} \Msun$, and can grow via two growth channels: BH mergers and gas accretion. Gas accretion is Eddington-limited and allowed to accrete at the Bondi-Hoyle-Lyttleton accretion rate. To define high accretion and low accretion states, it is checked whether the Bondi-Hoyle-Lyttleton accretion rate  exceeds a specific fraction of the Eddington limit by
\begin{equation}
    \mdotbondi/\mdotedd \geq f_{\rm Edd, th},
\end{equation}
where the Bondi-Hoyle-Lyttleton accretion rate and the Eddington limit are defined, respectively as
\begin{equation}
      \mdotbondi =\frac{4\pi G^2 M_{\rm BH}^2\rho}{c_{s}^3}  \quad\text{and}\quad
       \mdotedd =\frac{4\pi G M_{\rm BH} m_{\rm p}}{\epsilon_{\rm r}\sigma_{\rm T} c},
\label{eq:eddington}
\end{equation}
where $c$ is the speed of light in vacuum, $m_{\rm p}$ the mass of the proton, and $\sigma_T$ the Thompson cross-section. $\epsilon_{\rm r}$ represents the radiative accretion efficiency. $M_{\rm BH}$ is the mass of the black hole, while $\rho$ is the gas density of the surrounding gas; $c_{\rm s}$ is the effective speed of sound near the black hole and corresponds to $(c_{\rm s,therm}^2+B^2/4\pi\rho)^{1/2}$. 
The $f_{\rm Edd,th}$ scales with black hole mass to promote the transition from high to low accretion rates for the most massive black holes at late times, such that
\begin{equation}
    f_{\rm Edd.th} ={\rm min}\left[ f_{\rm Edd,0}\left( \frac{M_{\rm BH}}{10^8\Msun}  \right)^\beta, 0.1 \right],
    \label{eq:fedd}
\end{equation}
where $\beta$ and $f_{\rm Edd,0}$ are both free parameters.  It is worth noting that the values of $f_{\rm Edd,0}$, $\beta$, and the threshold mass $10^8\Msun$ do not have independent values. These values were selected so that the most massive black holes can attain the low accretion rate state while the less massive ones struggle to do so. 

In the model, there are two modes of AGN feedback, and the transition between the modes is given by Eq.~\ref{eq:fedd}. The two modes are (i) thermal \textit{quasar mode}  that heats the surrounding gas of the BH at high accretion rates \citep[e.g.][]{springel2005,diMatteo2005}. The feedback energy in the \textit{quasar mode} is released continuously as thermal energy that is injected into the surrounding gas given by
\begin{equation}
\Delta E_{\rm high}= \epsilon_{f,\rm high}\epsilon_r \dot{M}c^2\Delta t,
\label{eq:ehigh}
\end{equation}
where $\epsilon_r$ is the radiative efficiency, taking typical values between $0.04-0.4$, for optically thin, geometrically thin, and radiatively efficient accretion discs depending on the SMBH spin. In the original TNG runs, it has been set to $0.2$, and we do not change it to prevent additional degeneracy. The parameter $\epsilon_{f,\rm high}$ is the fraction of energy in \textit{quasar mode}  that couples with the surrounding gas, set to be $0.1$ in the original TNG simulation. (ii) The second mode is the kinetic \textit{wind mode} that produces winds, typically when the SMBH is massive enough and is subject to low accretion rates.

For our suite of zoom-in simulations, we consider two variations in the quasar mode feedback: (1) the model \textbf{NBH}, which is a simulation performed without BH physics, and (2) \textbf{BHlowEff}, where the energy fraction that couples to the surrounding gas in the AGN feedback in quasar mode,  $\epsilon_{f,\rm high}$, is set to a lower value or 0.05. Both models are compared to the TNG50-like model, which takes the same parameters as the parent simulation  \TNGF. The values adopted for the modified parameters are found in Table~\ref{table:models}.

\subsection{Kinematic decomposition and identification of bars}
To identify the different morphological components of each galaxy, we employ the kinematic decomposition algorithm \textsc{mordor} \citep{zana2022} to determine more specific galaxy components. The decomposition is based on the circularity ($\epsilon$) and binding energy ($E$) phase space, where a minimum in $E$ is identified.  The methodology identified five components: classical bulge, pseudobulge, thin disc, thick disc and stellar halo. To calculate the disc size, we use the face-on stellar surface density profiles of the total disc (thin and thick components), fitting an exponential profile.

A bar is identified using the Fourier decomposition of the face-on stellar surface profile. We use the maximum value reached in the ratio between the second and zero terms of the Fourier expansion ($A_2(R)$) as a proxy for the bar strength. The length of the bar is defined as the maximum radius where the phase ($\Phi(R)$) is constant. If the bar strength exceeds $0.2$, the galaxy is classified as barred. Appendix \ref{app:bars} provides detailed information and is based on \cite{rosasguevara2022}. The thickness of the disc (using both thin and thick components) was calculated by fitting the edge-on surface density of the disc component with a square arc secant function in terms of the z-direction. In the case of the bulge size,  we employ the face-on stellar surface density profiles, which are simultaneously fitted to the sum of a Sersic and an exponential profile for unbarred galaxies and the sum of two Sersic and an exponential profile where a bar is identified using the Fourier decomposition of the face-on stellar surface profile. Examples and more details of kinematic decomposition, identification of bars and the method of three-component decomposition of surface face-on density profiles can be found in Appendix \ref{app:decomp}. Table~\ref{table:zooms} shows the median values of the bulge and disc structural properties for the different models. It also includes the median values of the bar strength and length for the models. We do not find any particular trend for the bar strength when we vary models since the relation of the bar length with stellar mass is flat, indicating no direct dependence on stellar mass. From hereafter, we refer to a disc as the sum of the thin and thick disc components and a bulge as a classical bulge.  

\section{Impact on bar and galaxy properties}
\label{sec:galprop}
\begingroup
\renewcommand{\arraystretch}{1.5}
\begin{table*}
\begin{minipage}{\textwidth}
\caption{Median galaxy properties at $z=0$ for the different models. The methodology used to calculate disc, bar and bulge properties is provided in Appendix A. From left to right: Variation Model, the logarithm of halo mass, disc (thin and thick components) to total mass fraction from kinematic decomposition (from fitting the total surface density stellar profiles), bulge (classical component) to total mass fraction from kinematic decomposition (from fitting surface density stellar profiles), the effective radius from the bulge calculated from a Sersic profile, disc scale length from kinematic decomposition $^2$, disc thickness fitting the edge-on surface density with a square arc secant function in the z-direction, bar length and strength using Fourier decomposing face-on stellar surface density.}
\begin{tabular}{ccccccccccc} 
\hline
Sim Name  &  ${\rm log_{10}}(M_{\rm crit200}/M_{\odot})$     &   ${\rm log_{10}}(M_{*}/M_{\odot})$  & $D/T$ ($(D/T)_{\rm fit}$)  & $B/T$ ($(B/T)_{\rm fit}$) & $R_{\rm bulge}$  & $h_{\rm disc}$ & $h_{z, \rm disc}$  & $r_{\rm bar}$ & $A_{2,\rm max}$ \\

           &    &    & & & $\kpc$  &$\kpc$ & $\kpc$ & $\kpc$ &\\
\hline
\hline
NW (no winds)          &  $11.98$ &  $11.06$     &    $0.48$ 	 ($0.45$) &	 $0.21$ 	 ($0.55$)  &  $0.44$  &	  $1.33$ &  $0.55$  &	 -- 	& --
  \\
WW (weaker winds)      &  $11.91$ &  $10.70$     &    $0.51$ 	 ($0.59$) &	 $0.23$ 	 ($0.35$)  & $0.33$   &    $3.01$ 
& $1.58$ & $2.35$ &	 $0.62$  \\  
TNG50-like      &  $11.95$ &  $10.72$     &    $0.74$ 	 ($0.70$) &	 $0.10$ 	 ($0.18$)  &  $0.64$ &	 	 $2.98$ 
& $2.01$ &	 $2.45$ & 	 $0.50$ \\   
MW (medium winds)      &  $11.94$ &  $10.53$     &    $0.74$ 	 ($0.88$) &	 $0.08$ 	 ($0.11$)  &  $1.03$ &	  $5.19$ 
& $1.71$ &	 -- &	 --
  \\   
SW (strong winds)      &  $11.93$ &  $10.43$     &    $0.67$ 	 ($0.86$) &	 $0.09$ 	 ($0.10$)  &  $1.04$ &		 $4.47$ 
& $2.00$ &	-- & 	 --
 \\
NBH (no black holes)   &  $11.94$ &  $10.62$     &   $0.72$ 	 ($0.74$) &	 $0.10$ 	 ($0.20$)  &  $0.55$ &	  	$3.41$ 
& $1.66$ &	 $2.70$ &	 $0.57$ \\
BHlowEff              &  $11.92$ &  $10.58$      &   $0.63$ 	 ($0.78$) &	 $0.12$ 	 ($0.16$)  &  $0.79$ &		 $4.28$ 
& $1.80$  &	 $2.95$ &	 $0.53$ \\
\hline
\end{tabular}
\label{table:zooms}
\footnotesize{$^2$ The disc-scale length of galaxies in the no wind model is calculated using the stellar half-mass radius of the disc and assuming an exponential profile. We use this alternative because fitting an exponential profile of the disc stellar surface density fails due to the high density presented in the centre of these discs.}
\end{minipage}
\end{table*}
\endgroup


In this section, we investigate the impact of altering the strength of the winds and BH physics on the properties of the galaxies and their structural parameters as a function of redshift.

\subsection{Properties of bars and galaxies at $z=0$}
The median mass of the halo in the original TNG50 simulation is $M_{crit200}=10^{10.91}\Msun$ (see Table \ref{table:haloes}) and the median of the zoom-in haloes for the different models ranges between $M_{200}=10^{10.92}\Msun$ (BHlowEff model) and $M_{200}=10^{10.98}\Msun$ (NW model) at $z=0$ which is at most a difference of $0.06$ dex in halo mass. This is not the case for the rest of the properties when higher dispersion is found in varying supernovae feedback.  Table \ref{table:zooms} shows some of the median properties of the galaxies and their bars at $z=0$. The median stellar mass ranges between  $M_{*}=10^{10.43}\Msun$ (SW model) and $M_{*}=10^{11.06}\Msun$ (NW model) at $z=0$. When this is compared to the stellar masses of the halos in the original TNG50 simulations ($M_{*}=10^{10.61}\Msun$) or TNG50-like model ($M_{*}=10^{10.72}\Msun$), the difference varies between $0.4$ and $0.2$ dex in stellar mass. The most massive galaxies are those without SN feedback (NW model), whereas the least massive galaxies are those in the SW model. This is expected since supernova processes are known to be physical processes that eject gas or prevent gas from forming stars, and could influence the angular momentum of the gas in the galaxy.
It is interesting to note that the morphology roughly remains the same as in the original TNG50 simulations at $z=0$ in the sense that galaxies have a high disc dominant component, although the stellar mass and sizes of the disc and the bulge are different in all the models (see Table~\ref{table:zooms}). The galaxies with the lowest $D/T=0.48$ ($(D/T)_{\rm fit}=0.51$, using Fitting surface profiles) are those in the NW model, whereas the highest $D/T=0.74\,((D/T)_{\rm fit}=0.88)$ are those in the MW model. In the case of the bulges, the lowest $B/T=0.09$  ($(B/T)_{\rm fit}=0.10$) is for the galaxies in the SW model, whereas the highest $B/T=0.21$ is for the galaxies in the WW model (and in the NW model using $(B/T)_{\rm fit}=0.55$). The compactness also changes. The most compact and massive bulges and thinner and massive discs are those in galaxies in the NW model ($R_{\rm bulge}=0.44$ kpc, $h_{\rm disc}=1.33$ kpc, $h_{z}=0.55$ kpc). We note, however, that the smallest bulge radii correspond to galaxies in the weak wind model. The least compact bulge ($R_{\rm bulge}\approx 1$ kpc is found in the galaxies in SW and MW models. Also, the galaxies in SW and MW models exhibit the most extended and thicker discs found in galaxies with $h_{\rm disc}=4.47$ kpc, $h_{z}=2.00$ kpc and $h_{\rm disc}=5.19$ kpc, $h_{z}=1.71$ kpc, respectively.
Regarding the galaxy properties in the models varying BH physics, the galaxy properties are pretty similar to those in the  TNG50-like model.
The parent TNG50 galaxies exhibit a bar, while in the zoom-in simulations, $83\%$ of them present a bar at $z=0$ ($5$ of $6$ galaxies) in the TNG50-like model. This galaxy, in particular, does not form a bar in any of the models, including the weak wind model, where the bar is present in the other galaxies. The median bar sizes lie between $2$ kpc and $3$ kpc, and the disc sizes all of which are strong bars ($\Amax\geq 0.4$) as in the parent TNG50 simulations.
\begin{figure*}
\includegraphics[width=2\columnwidth]{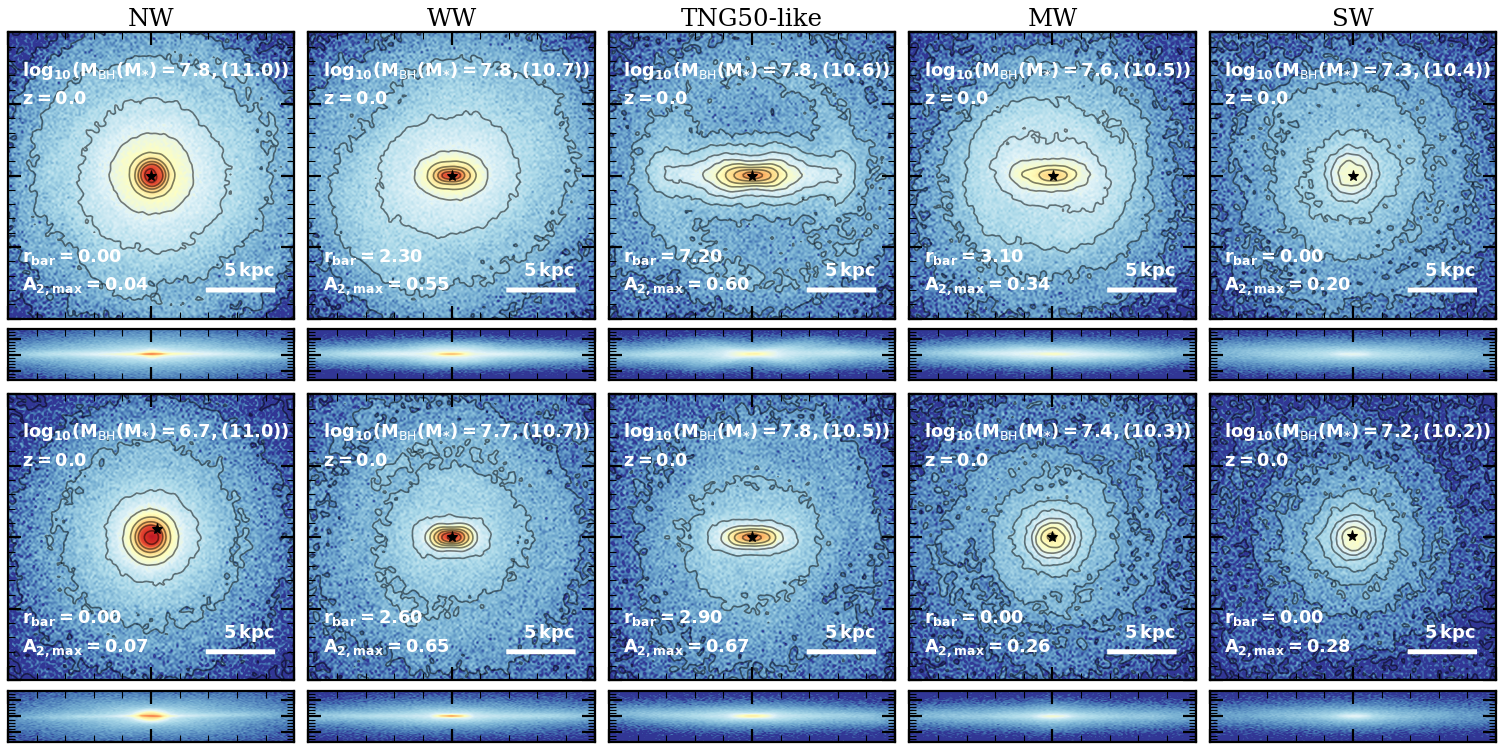}

\includegraphics[width=1.2\columnwidth]{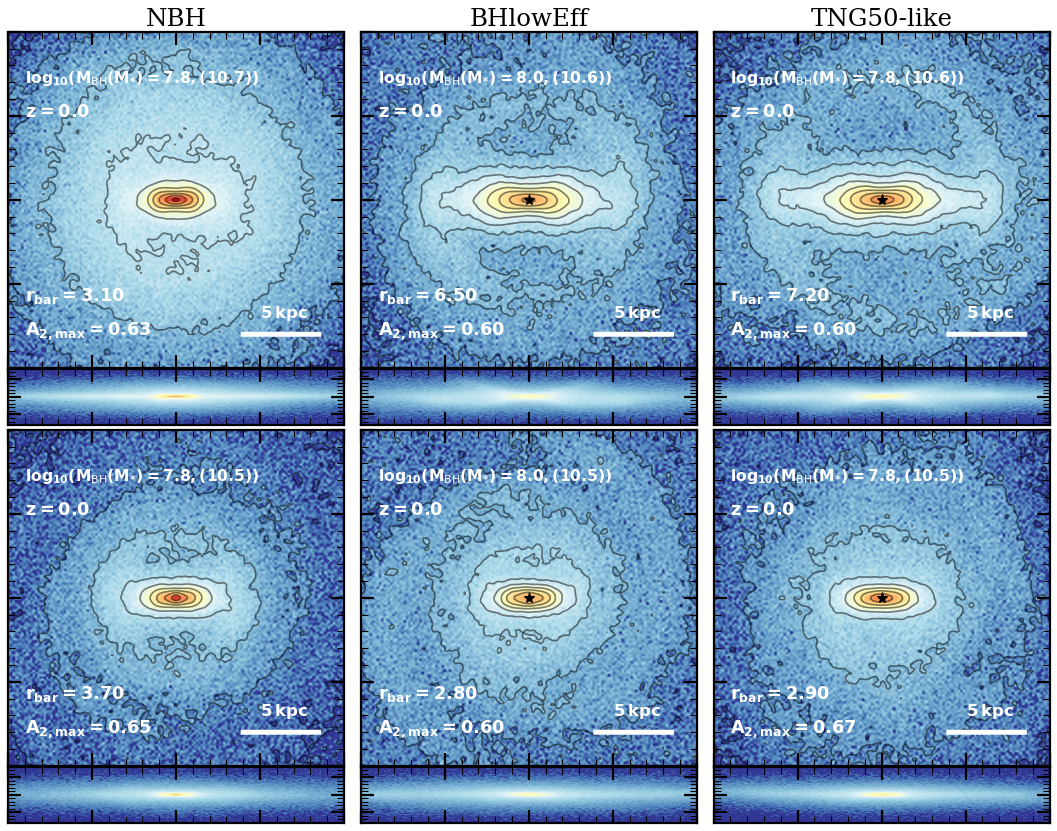}
\caption{Face-on stellar density maps of two distinct galaxies at $z=0$. Top figure shows the wind model variations: each column corresponds to a distinct wind model, with left Col. showing a galaxy without SN feedback and right Col. galaxies with the strongest SN feedback. The bottom figure shows different quasar black hole models but with the same stellar feedback model as TNG50-like as indicated by Table~\ref{table:models}. The black star symbolises the location of the black hole. In both the no wind and strong wind models, the disc galaxy does not exhibit the formation of a bar. Quasar black hole physics appears to have a limited influence on the formation of a bar.}

\label{fig:StellarDensityMaps}
\end{figure*}
The top panel of Fig.\ref{fig:StellarDensityMaps} illustrates two galaxies at $z=0$ that are subject to the various variations in the wind model from NW (no wind model) to SW (strong wind model). The face-on (top panels) and edge-on (bottom panels) stellar density maps are represented in each column, with the energy injection per SN event intensifying from left to right. The disc-like morphology of the galaxy is generally independent of the model, as mentioned previously.  The figure shows that the bar does not form in either the NW model or the SW model. The galaxy appears to be more concentrated in the no-wind model, while the galaxy appears to be less concentrated in the SW model. The two objects exhibit this behaviour. The bar forms in the TNG50-like model and the WW model. For the MW model, only one of the galaxies forms a bar. It is essential to acknowledge, however, that the stellar bar properties, including its length and strength, are different for different models.

The two objects are also depicted in the bottom panel of Fig.\ref{fig:StellarDensityMaps} in two distinct variations in the BH physics: a model without BH physics (NBH) and a BH with less efficient quasar AGN feedback (BHlowEff) when the BH is accreting at higher rates compared to the TNG50-like model but with the same stellar feedback assumed as in the TNG50-like model. We observe that galaxy morphology is not significantly altered by the variations in the BH model, as opposed to the variations in the wind models. However, the final properties of the bar and the evolution of the BH itself will differ.

\begin{figure*}
\begin{tabular}{c}
\includegraphics[width=1.9\columnwidth]{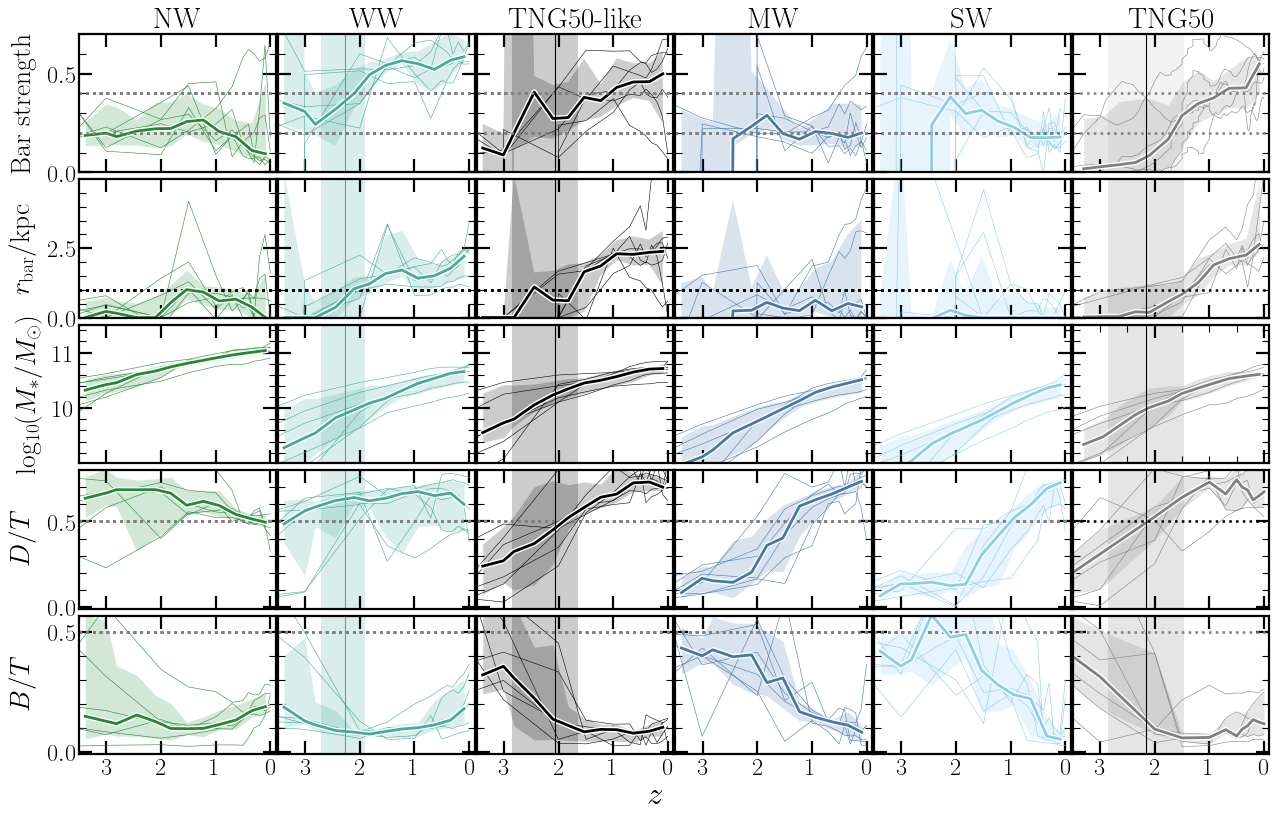}
\\
\includegraphics[width=1.3\columnwidth]{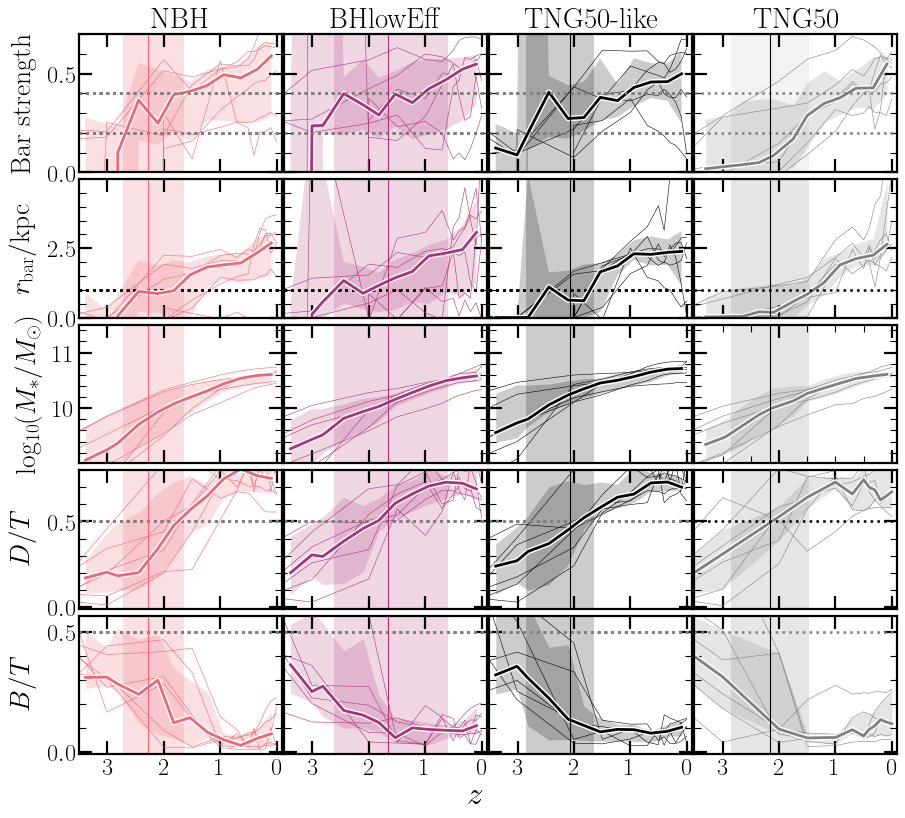}
\end{tabular}
\caption{ Median evolution of the bar and galaxy morphology for simulations with different galactic wind (top) and black hole feedback (bottom) models. Each column corresponds to a different model, as indicated in the figures. From top to bottom rows: the bar strength, bar extent, the stellar mass and the disc-to-total and bulge-to-total mass fraction are shown. The medians and the distribution between $20^{\rm th}$ and $80^{\rm th}$ percentiles correspond to solid lines and shaded regions, respectively. Thinner solid lines correspond to the evolution of individual galaxies. Vertical solid lines and shaded region represent the median redshift and the distribution between $20^{\rm th}$ and $80^{\rm th}$ percentiles of the bar formation time. The stronger the wind, the lower the probability of the galaxy developing a bar, except for the no-wind galaxy. The build-up of a well-defined massive cold disc is delayed for stronger galactic wind models. No significant change is seen in the black hole variation models.}

\label{fig:evolutionwinds}
\end{figure*}

\subsection{Galactic bar and galaxy evolution}
\begin{figure*}
\begin{tabular}{c}
\includegraphics[width=1.9\columnwidth]{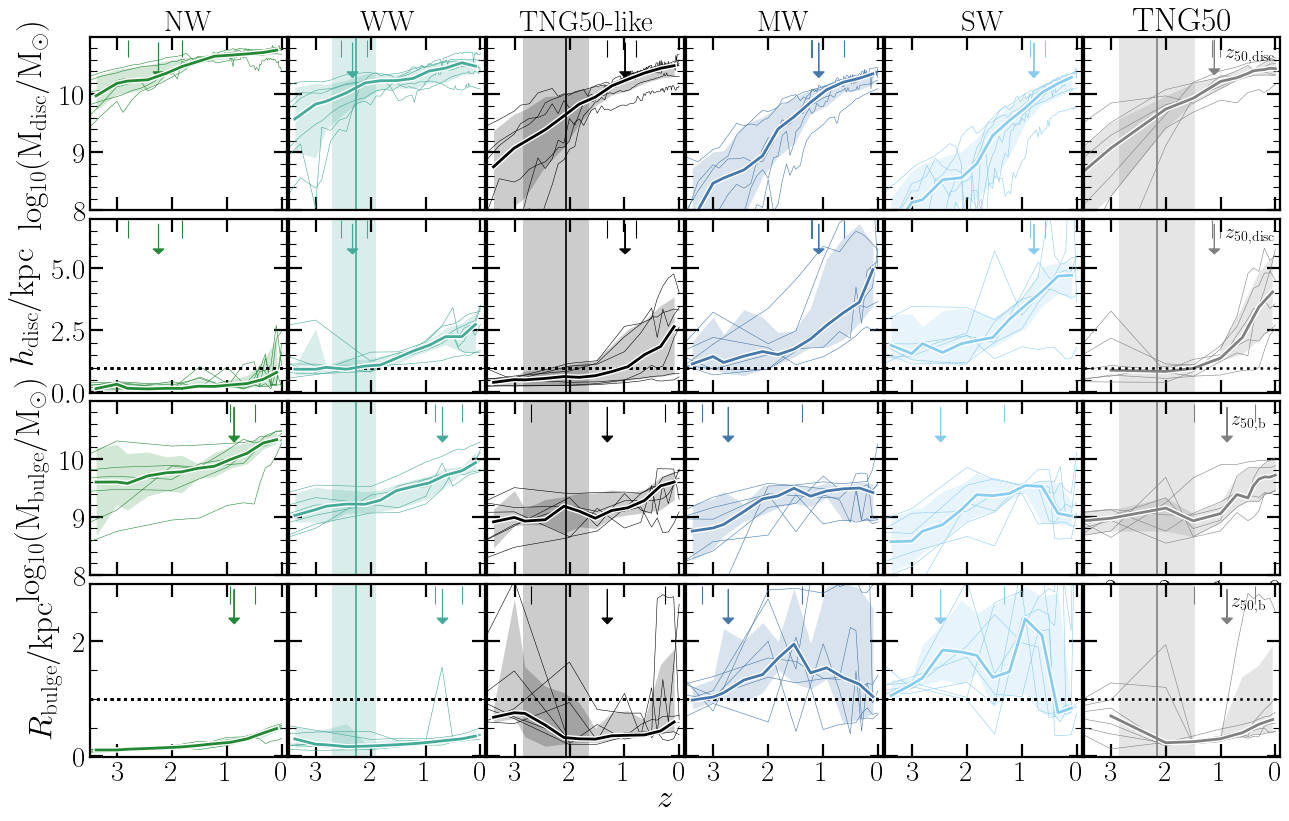} \\
\includegraphics[width=1.4\columnwidth]{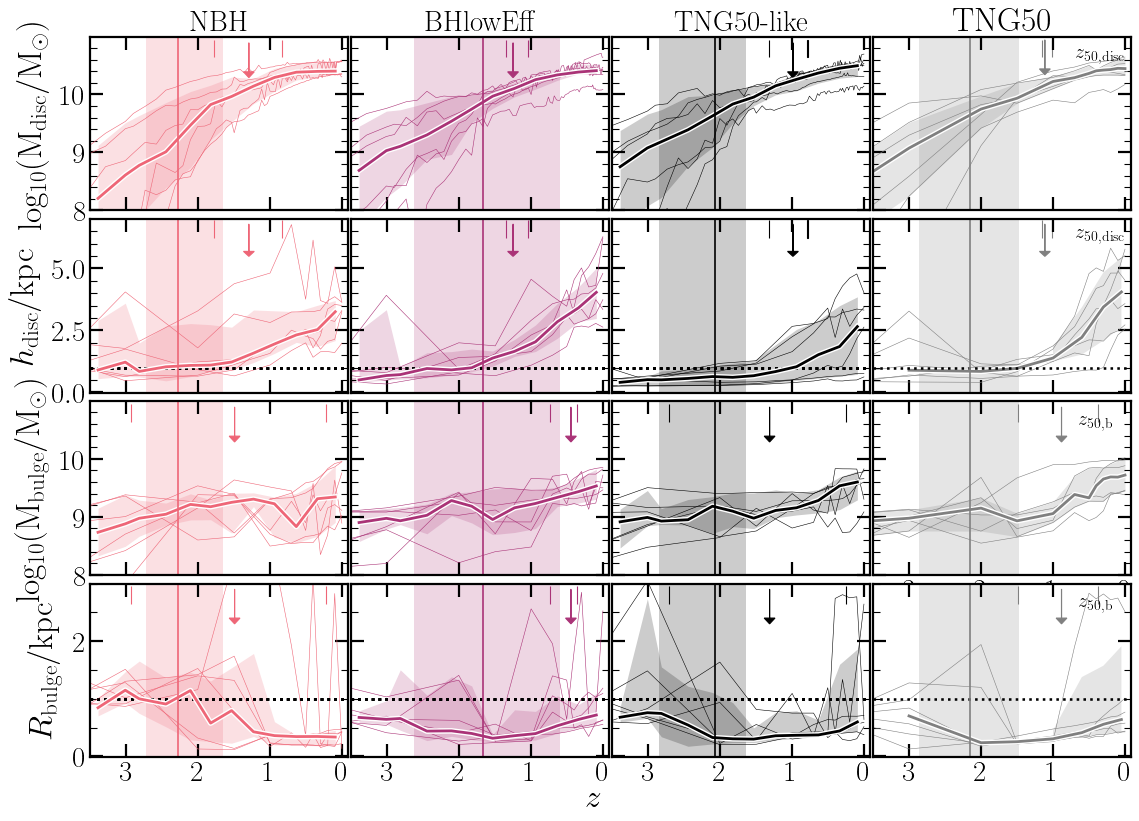}
\end{tabular}
\caption{ Mass and size evolution of discs and bulges.  Galactic winds model variations (top) and quasar BH physics model variations (bottom). The last Col. shows the original TNG50. Solid lines and shaded regions represent median values and the $20^{\rm th}$ to $80^{\rm th}$ percentile distribution.  Thinner lines correspond to the evolution of each galaxy. Vertical solid lines denote bar formation median redshift, and shaded regions its scatter.  Arrows with small vertical lines correspond to the median redshift formation of the disc/bulge ($z_{50,\rm disc}$/$z_{\rm b}$) and its scatter. Stronger SN feedback delays disc buildup and makes discs more extended. }
\label{fig:evolutioncomp}
\end{figure*}

We focus now on the evolution of the bar and galaxy properties.
Fig.~\ref{fig:evolutionwinds} illustrates the median redshift evolution of the bar properties and morphology of the galaxies for each of the four galactic winds models, the three BH models, and the galaxy counterparts in the TNG50 simulation.  We observed that only the galaxies in the WW and TNG50-like models develop a stable bar in the galactic wind models. The galaxies in the other variations do not form a stable bar. We also observe that the weak wind model develops the strongest bars, which are also shorter in comparison to the galaxies in the TN50-like model. It is important to note, however, that galaxies in the no wind model develop non-axisymmetric instabilities, but they do not grow over time. This suggests that the disc may be unstable (see Sect. \ref{sec:barformation}); however, refrain from forming a bar. 
We observed that the galactic bar formed earlier (see cyan vertical lines for the weak wind model,  $z_{\rm bf}=2.5$) compared to the TNG50-like models (black vertical lines, with $z_{\rm bf}=2$). However, the scatter in the time of bar formation is large in all the models (see vertical colour lines in Fig.~\ref{fig:evolutionwinds}), including TNG50 galaxies. This could be attributed to the chaotic-like behaviour caused by feedback processes in the TNG model, affecting the scatter in some galaxy properties \citep[e.g.][]{genel2019,joshi2024}.

The morphology of the galaxies in the NW and WW models has been well-defined since early times. For instance, the galaxies in the no wind model have already assembled a substantial disc by $z_{\rm disc}=3$ and are also the most massive galaxies in comparison to the other models. When we move to later times, the massive disc component is delayed when the wind intensity generated by SN feedback is greater, whereas the bulge forms earlier and is less massive in comparison to the disc. It should be noted that the bar formation seems to be linked to the earlier build-up of the massive disc (see \citealt{zana2019,rosasguevara2020,izquierdo2022,khoperskov2024}). However, galaxies do not form a stable bar in the no wind model. Conversely, galaxies in the strong wind models developed a substantial disc that was later formed, but it never developed a stable bar. In the BH variations models, the bar is present in all simulations, but the bar formation time varies dramatically. Additionally, we can observe some variations in the length and strength of the bars for galaxies across the various models. The BHlowEff variation, in particular, displays the bars with the highest length and strength.

Fig.~\ref{fig:evolutioncomp} illustrates the evolution of the disc and bulge masses and their sizes in the galactic wind and the quasar BH physics models. The galaxies in all models possess a massive disc which is larger than $10^{10}\Msun$ and becomes dominant, but their build-up is distinct: the disc build-up is slower as the winds from supernova feedback become stronger. In order to quantify this, we calculate the formation time (redshift) of the massive disc by determining the time at which it reaches $50\%$ of its $z=0$ mass. This is depicted as vertical dotted lines in the figures, with the 20th and 80th percentiles for all objects. These measured times (redshifts), $t_{50,\rm disc} (z_{50,\rm disc})$, show that the disc formation occurs later for galaxies in the strong wind model ($z_{50,\rm disc}=0.8$) and earlier for galaxies in the no wind model ($z_{50,\rm disc}=2.25$). We note that the scatter of $z_{50,\rm disc}$ in the variations of the BH model is higher than that in the TN50-like model, oscillating between $z=0.8$ and $z=1.3$ in the case of galaxies in the no black holes model. However, the growth of the disc is not significantly affected.

The second row of Fig.~\ref{fig:evolutioncomp} illustrates the disc length. This generally increases with increasing time for all the models; however, the final disc length increases as the intensity of the galactic winds increases. The most extended and least dense discs are those in the strong wind model, while the most compact and dense are in the no wind model. This is important because these properties play a role in the bar formation criterion, as we will see in the next section.

The bulge growth for variations in the wind and BH models is illustrated in the third row of Fig.~\ref{fig:evolutioncomp}. The figures demonstrate that the bulge evolution is more complicated, particularly in light of the wind model variations. Additionally, it is important to observe that the fluctuations between galaxies in each model are greater. In the case of NW and WW models and TNG50 galaxies, the bulge forms later in time than the disc, suggesting that the stellar bar may facilitate the expansion of the bulges.

The SN feedback intensity is high in models with moderate (MW) and (SW) strong winds, resulting in a more complex bulge assembly, subject to significant fluctuations. Initially, the bulge experiences an increase in mass as time progresses, reaching up to $50\%$ of its current mass even earlier. Subsequently, the bulge mass decreases over time. We track the stellar particles that are part of the bulge at the moment of their mass peak. We find that a fraction of the stellar particles that were not part of the bulge later in time present higher velocities and at larger distances, associated with the halo component. This might be due to a fast ejection of matter.

The bulge size is presented in the bottom panels of the figure. As the bulge mass, their evolution is more complicated, and it is more affected by the wind model variations. The bulge size increases with time for NW, WW, and TNG50-like models; the MW and SW models are more complicated and show similar fluctuations, as seen in the bulge mass attributed to the intensity of SN feedback. The most compact and massive bulges are those in the no wind model. The most extended and least massive bulges are found in the strong wind models.

In the strongly barred galaxies, discs and bulges are formed at a more rapid pace than those in unbarred galaxies. Our results suggest that gas is swiftly consumed to form stars in galaxies in the models with NW, WW, and TNG50-like models, resulting in more dense discs and bulges.  In contrast, galaxies in models with medium and strong feedback discharge gas at earlier times as a result of the less intense gravitational potential of the halo and the intense SN feedback. 
This would have an impact on the properties of the gas and the formation of stars from early times. This is in agreement with the result from \cite{bi2022a}. The authors, using zoom-in simulations with a mass resolution compatible with the TNG50 simulation, find that SN feedback impacts the cold accretion flow (see their Figure 1) in high-redshift Milky-Way galaxies.

These galaxies properties have been used to assess the stability of a disc against gravitational instabilities, such as bars, which we will study in the next section.

\section{Bar formation}
\label{sec:barformation}

We study whether the galaxies with different galactic winds and quasar BH physics model variations satisfy bar instability criteria, which are contingent upon the structural parameters. In particular, we will be focussing on three criteria: the \citealt{toomre1964} (Eq.\ref{eq:Toomre}), \citealt[][(ELN)]{ efstathiou1982} (Eq.\ref{eq:ELN}), and \citealt[][(MMW)]{mo1998} (Eq. \ref{eq:MMW}) criteria.

\begin{figure*}
\begin{tabular}{c}
\includegraphics[width=2\columnwidth]{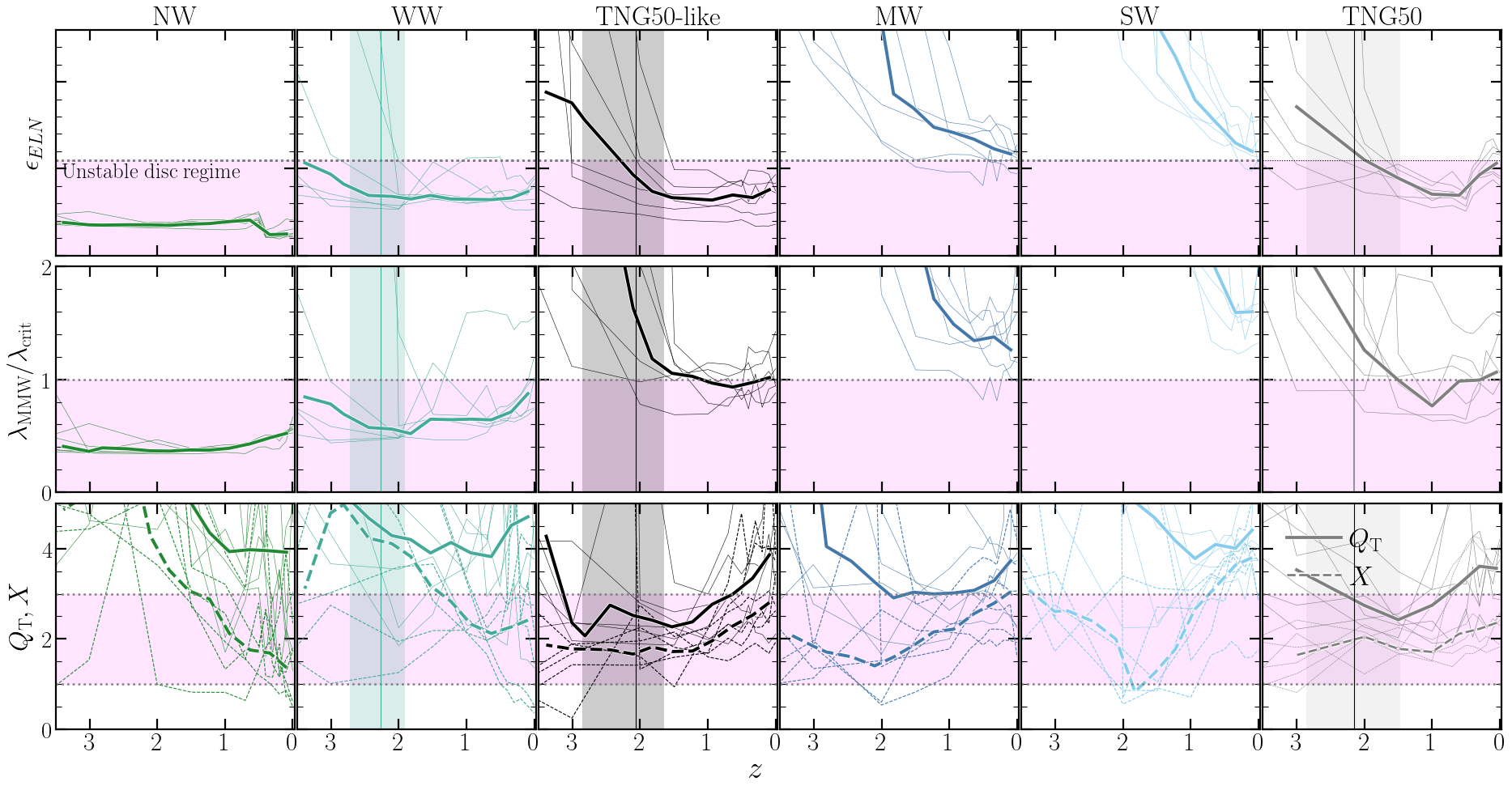}

\\
\includegraphics[width=1.4\columnwidth]{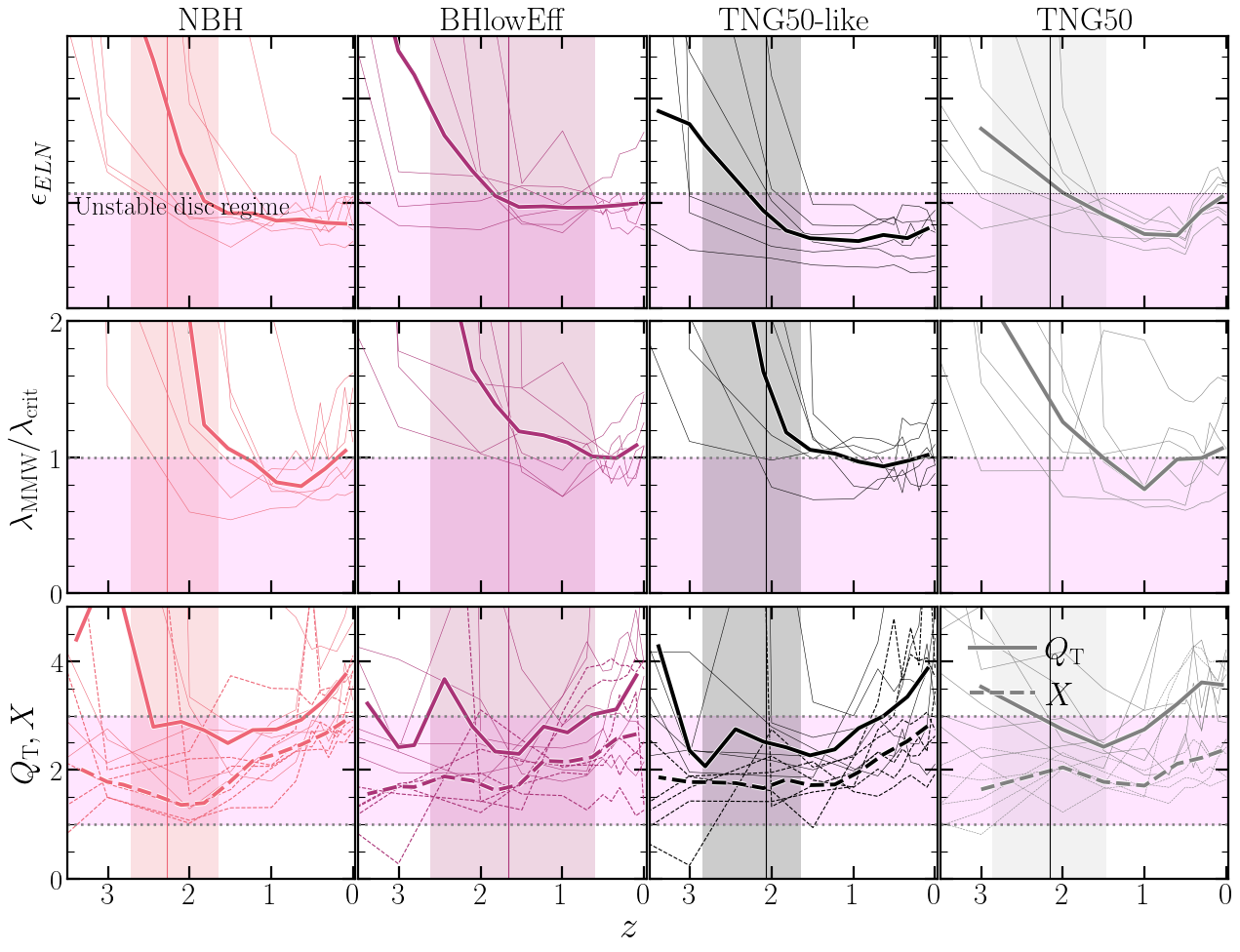}
\end{tabular}
\caption{Evolution of relevant parameters for bar formation. From top to bottom: ELN criterion (Eq.\ref{eq:ELN}), MMW criterion (Eq.\ref{eq:MMW}) and Toomre parameter (Eq. \ref{eq:Toomre}) for different galactic wind models (top figure) and BH physics models (bottom figure) and the original TNG50. Horizontal dashed lines represent the threshold that defines if a disc is unstable, and magenta shaded region represents the values where the disc is unstable. Thinner lines correspond to the evolution of each halo. Vertical solid lines and shaded region represent the median redshift and the distribution between $20^{\rm th}$ and $80^{\rm th}$ percentiles of the bar formation time. For TNG50-like, MW and SW models, the three instability criteria determine if a bar forms or not. For NW  model, ELN and MMW criteria fail to predict a stable disc, whereas Toomre criterion fails to predict bar formation for WW model. }
\label{fig:ToomreELNLambda}
\end{figure*}

The evolution of the parameters  $\epsilon_{\rm ELN}$,  $\lambda_{\rm MMW}$ and $Q_{T}$ is illustrated in Fig. \ref{fig:ToomreELNLambda} for the galactic wind models (top figure) and BH physics models (bottom figure) models. Except for the no wind model, the  parameters $\epsilon_{\rm ELN}$,  $\lambda_{\rm MMW}$ roughly determine whether the disc will be stable to bar formation.
In particular, for WW and TNG50-like models, both criteria have been satisfied since the time of bar formation, since their values lie in the magenta-shaded region below the horizontal dotted line in the figure. In the case of the Toomre criterion, the Toomre parameters present a higher scatter. However,  the parameter $Q_{\rm T}$ is roughly able to determine if the galaxies are stable to bar formation. The exceptions are galaxies in the no wind and weak wind models.


\subsection{ ELN criterion}
\begin{figure*}
\begin{tabular}{c}
\includegraphics[width=2\columnwidth]{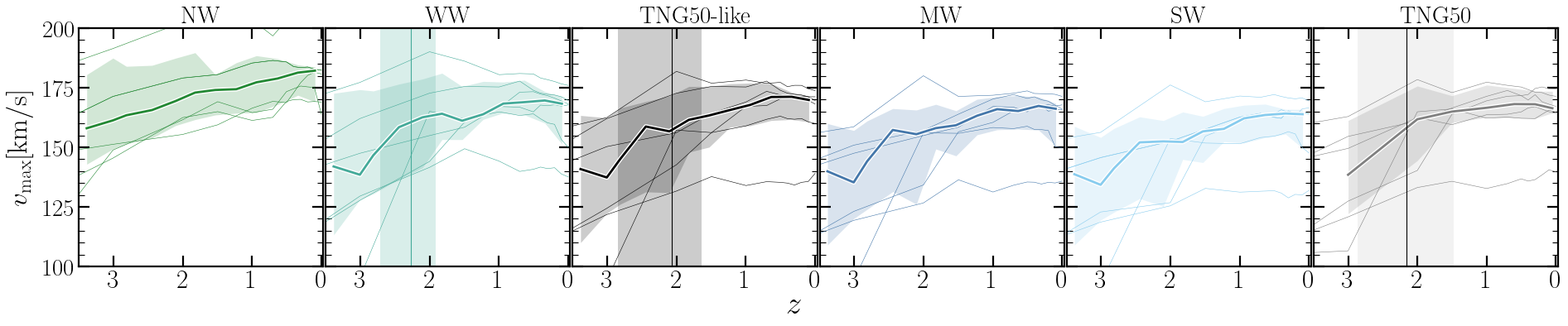} \\
\includegraphics[width=1.4\columnwidth]{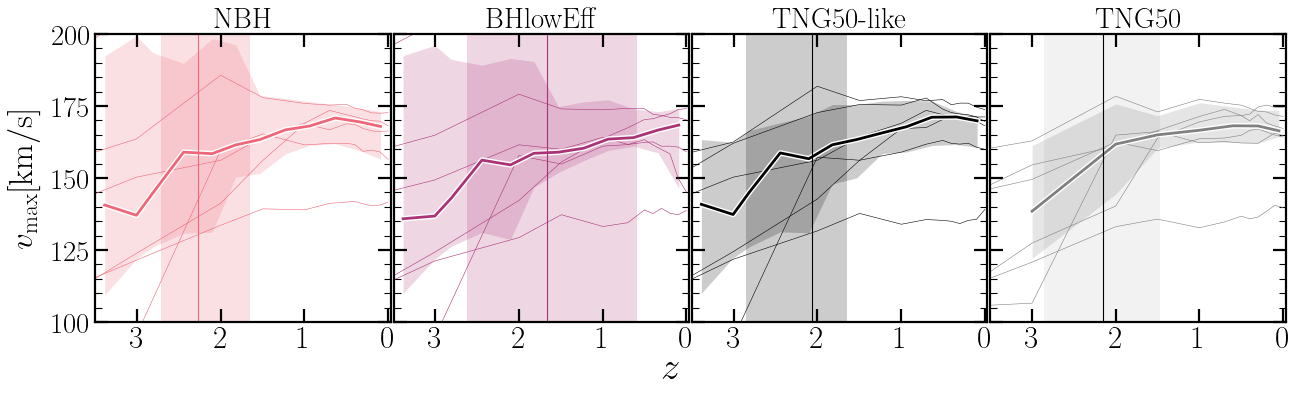}
\end{tabular}
\caption{Median evolution of the maximum circular velocity of the Dark Matter halo. Each column represents the variations of galactic winds (upper Fig.) and BH physics models (bottom Fig.) in order from left to right, and vertical solid lines represent the bar formation redshift. The original galaxy in TNG50 is represented by the final right Col. Thinner lines correspond to the evolution of each halo. Vertical solid lines and shaded region represent the median redshift and the distribution between $20^{\rm th}$ and $80^{\rm th}$ percentiles of the bar formation time. No significant variation is found in the evolution of $v_{\rm max}$ among the different models. }
\label{fig:ELNcriterion}
\end{figure*}

In this section, we focus on the evolution of the disc mass and size evolution and the maximum circular velocity ($V_{\rm max}$) of the dark matter halo that concerns the ELN criterion \citealt[][(ELN)]{efstathiou1982}. In the last section, in Fig. \ref{fig:evolutioncomp}, we have studied the mass and size growth of the disc for the different galactic wind and BH model variations. We observe that the evolution of the disc scale length and the disc mass have been affected by the strength of the galactic winds (kinetic energy released into the ISM) since early times. The most extreme case is the variation where there is no galactic wind or, opposite, the galaxy experiences a strong galactic wind. In both NW and SW models, the galaxy forms a massive disc ($M_{*}>10^{10}\Msun$) and a relatively small bulge ($B/T<0.4$, see Table \ref{table:zooms}), and neither of both models forms a stable bar. The other models show an intermediate effect between these two examples, and only the galaxies in the TNG50-like and weak wind models form a bar. In contrast, the evolution of the maximum circular velocity ($V_{\rm max}$) of the dark matter halo almost remains constant across all scenarios, as seen in Fig. \ref{fig:ELNcriterion}. There is no distinction when the dark matter halos are the sole factor considered. Therefore, the size of the disc is a determinant of the ELN criterion and, with less significance,  the stellar mass disc.
The ELN criterion fails to predict the disc stability of the galaxies in the no wind model, while the ELN criterion is met by the galaxy in the  WW and TNG50-like models, at least during the time of bar formation.

The BH physics variations show some differences in disc size evolution, but not in disc mass. However, the values and evolution of the $\epsilon_{\rm ELN}$ parameter remain consistent, as seen in Fig. \ref{fig:ToomreELNLambda}. The ELN criterion emphasises that a compact, dense disc is a prerequisite for forming a bar; however, our findings show this condition alone is not sufficient.



\subsection{MMW criterion}

In order to conduct a more thorough investigation of the MMW criterion \citep[][]{mo1998}, the middle panels of Fig.~\ref{fig:ToomreELNLambda} show the values of the parameter $\lambda_{\rm MMW}$, and the magenta shaded region corresponds to the values of $ \lambda_{\rm crit}$
as seen in Eq. \ref{eq:MMW}. For details in the calculation of $ \lambda_{\rm crit}$, see Eqs. 23, 32, 34 in \citet[][]{mo1998}. Therefore  if $(\lambda_{\rm MMW}/ \lambda_{\rm crit})>1$, the disc is approximately stable. For the galaxies in the models that form a bar, the  MMW criterion correctly determines that the disc is prone to form a bar since the parameter $(\lambda_{\rm MMW}/ \lambda_{\rm crit})\lsim 1$\footnote{We used the assumption that the halo concentration $c_{NFW}=10$ in order to compute $\lambda_{\rm crit}$; however, we experimented with smaller/larger values and got comparable results. This is because the criterion weakly depends on concentration as seen in \cite[][]{mo1998}.} lies in the magenta-shaded region, especially at the time of bar formation. An exception are the galaxies in the models without SN feedback, where the MMW criterion fails to predict that the disc is stable since a bar does not form. On the other hand, the MMW criterion is satisfied for galaxies in the models with medium and strong winds since they do not form a bar and have $(\lambda_{\rm MMW}/\lambda_{\rm crit})>1$. 

\begin{figure*}
\begin{tabular}{c}

\includegraphics[width=2\columnwidth]{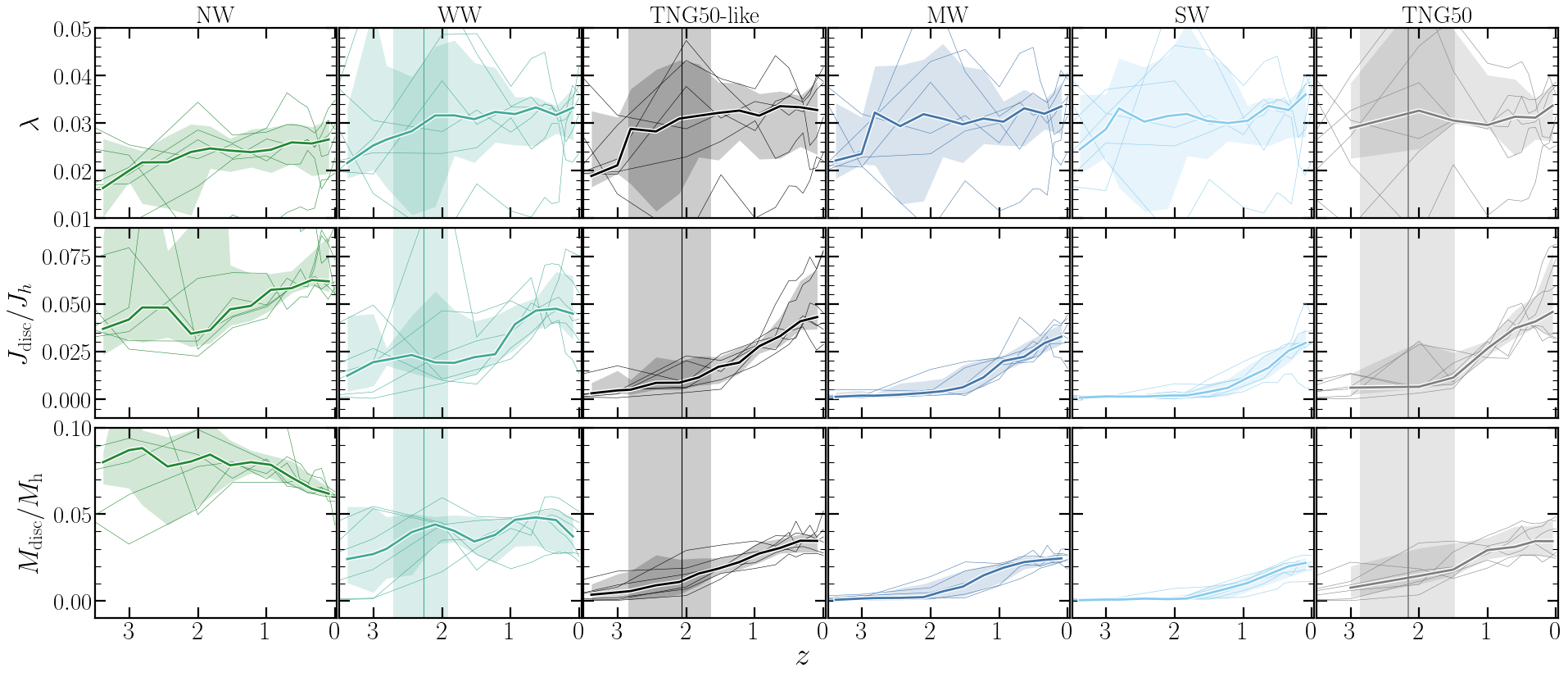}  \\
\includegraphics[width=1.3\columnwidth]{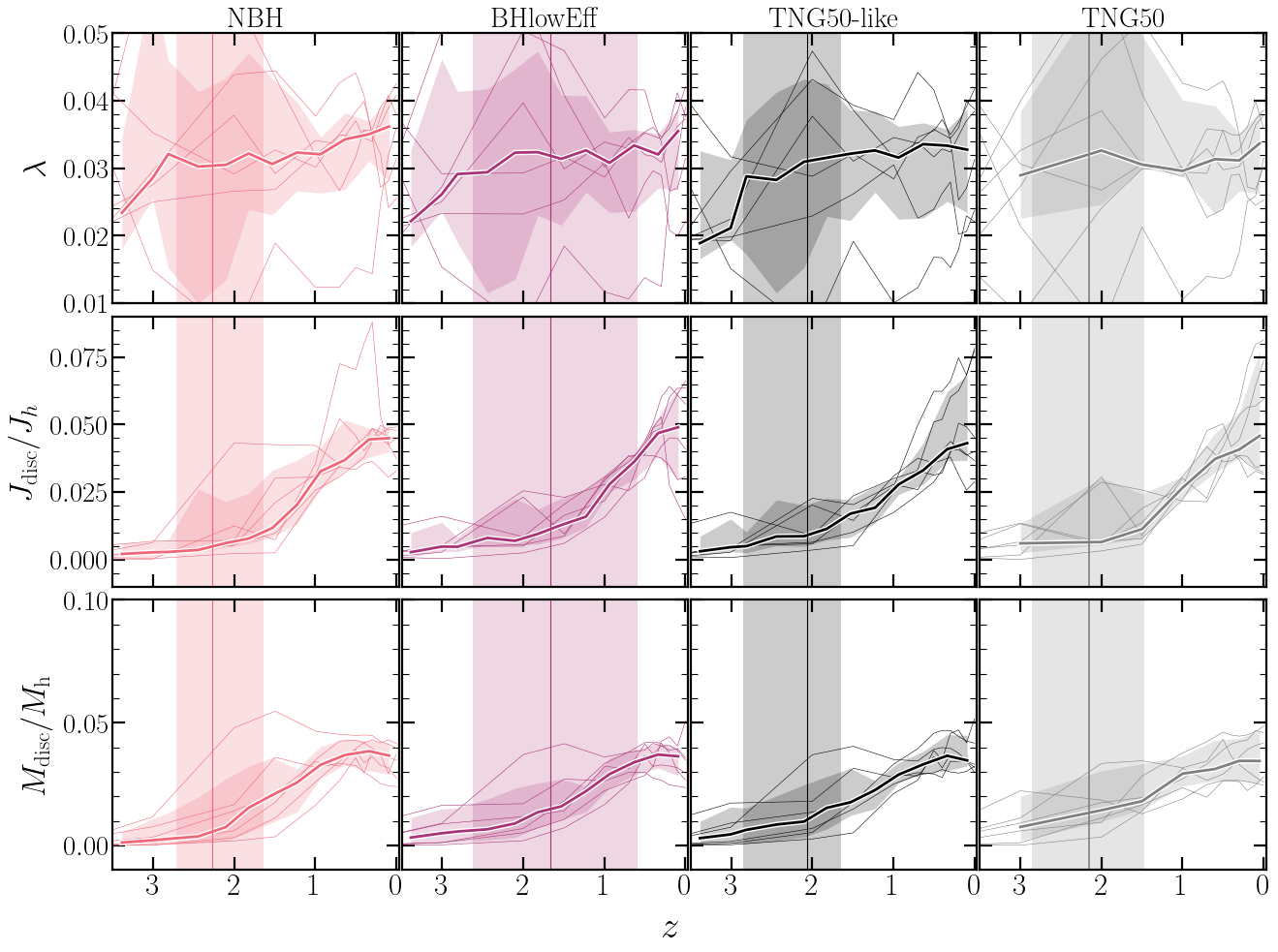}
\end{tabular}
\caption{Median evolution of spin parameter, the ratio of the angular momentum of the disc over the halo and the disc mass fraction (solid thicker lines). From left to right, different galactic wind models (top Fig.) and BH physics models (bottom Fig.) are shown. The rightmost Col. corresponds to the original galaxy in TNG50. Thinner solid lines correspond to individual halos. Vertical solid lines and shaded regions represent the median redshift and the distribution between $20^{\rm th}$ and $80^{\rm th}$ percentiles of the bar formation time. The angular momentum and mass of the disc with respect to the halo are smaller with stronger SN feedback. As the ELN criterion, the MMW criterion demonstrates the essential condition of the early buildup of a massive disc to favour bar formation. However, it is insufficient as it fails to predict stable discs in the NW model.}
\label{fig:criterionMo}
\end{figure*}

To understand why the criterion fails in some cases and others not, we follow the evolution of the galaxy properties that the parameter $\lambda_{\rm MMW}$ depends on: the spin parameter ($\lambda$), the ratio of the angular momentum of the disc and the halo ($J_{\rm disc}/J_{\rm h}$), and the ratio of the disc mass and the dm halo mass ($M_{\rm disc}/M_{h}$). The evolution of these galaxy properties is presented in Fig.\ref{fig:criterionMo}. The evolution of the spin parameter remains almost constant regardless of the galactic wind model or the varied BH physics, with values between $0.02$ and $0.04$, with the exception of the no wind model.  This model presents spin parameters smaller than $0.03$. On average,  the spin parameter in all the models presents low values in agreement with those studied in N-body simulations in favour of the formation of a bar \citep[e.g.][]{long2014}. We observe the spin parameter to be almost unaffected by SN feedback and BH physics model variations. However, $J_{\rm disc}/J_{\rm h}$ is subject to change as the intensity of the galactic winds is modified. This ratio increases over time, transitioning from values less than $0.02$ at $z=2$ to values just smaller than $0.05$ at $z=0$, except for the galaxies in the no wind model. These galaxies present high ratios and reach the highest value ($J_{\rm disc}/J_{\rm h}\sim 0.06$ at $z=0$).  The galaxies with the lowest values are those in the medium and strong wind models,  nearly reaching $0.04$ at $z=0$. This is expected since SN feedback allows for gradual gas cooling and accretion from a hot and diffuse gas halo that, in some cases, galaxies might form stable discs \citep[e.g.][]{sales2012}. Similarly, $M_{\rm disc}/M_{h}$ is affected by SN feedback with the highest values presented for the no wind model ($\gsim 0.05$) and the lowest values for the strong wind models ($\lsim 0.02$) at $z=0$.  Apart from the no-wind model, this ratio increases with time. Strong feedback can expel baryonic matter from galaxies since early times, cycling it through a hot and diffuse gas before it can be re-accreted and converted into stars \citep[e.g.][]{sales2012}. We note that for the WW, TNG50-like models, where galaxies form a bar, present higher values of high angular momentum and higher disc content with respect to the halo.  This result agrees with the ELN criterion that a massive disc is a necessary condition for a disc to form a bar, but not sufficient.  A good example is the galaxies in the NW model. In this case, the parameter $(\lambda_{\rm MMW}/\lambda_{\rm crit})$ determines that the discs should form a bar that is not there. As we will see in Sect. \ref{sub:bulgeeffect}, both the ELN and MMW criteria do not take into account the effect of the presence of a large and compact bulge.

\subsection{Toomre Criterion and Swing parameter}
As discussed in the introduction, the \cite[][$Q_{\rm T}$, see eq.\ref{eq:Toomre}]{toomre1964} is one of the criteria for bar instability. This parameter is responsible for disc stability in axisymmetric density perturbations. If $Q_{\rm T}<1$,  the disc is unstable. In conjunction with the Toomre parameter, the swing parameter ($X$) evaluates the amplification of the density perturbations through the swing amplification effect. This parameter is defined by  $X=r\kappa^2/(4\pi G) \Sigma$, where $\kappa$ is the epicyclic frequency, $G$ is the gravitational constant, and $\Sigma$ is the stellar surface density. Both parameters $Q_{\rm T}$ and $X$ depend on the kinematic properties of the disc as a function of the radius. The Toomre parameter should be approximately $1$, but it should be at most $2$ to achieve substantial amplification \citep[e.g.][]{binney2008}. To the extent that $Q_{\rm T}$ is excessively high, the disc is too stable to generate large-scale perturbations. 
In addition,  a value in the range  $1\leq X\leq3$ suggests a regime in which swing amplification can substantially amplify non-axisymmetric perturbations, such as spirals or bars.

\begin{figure*}
\begin{tabular}{c}

\includegraphics[width=2\columnwidth]{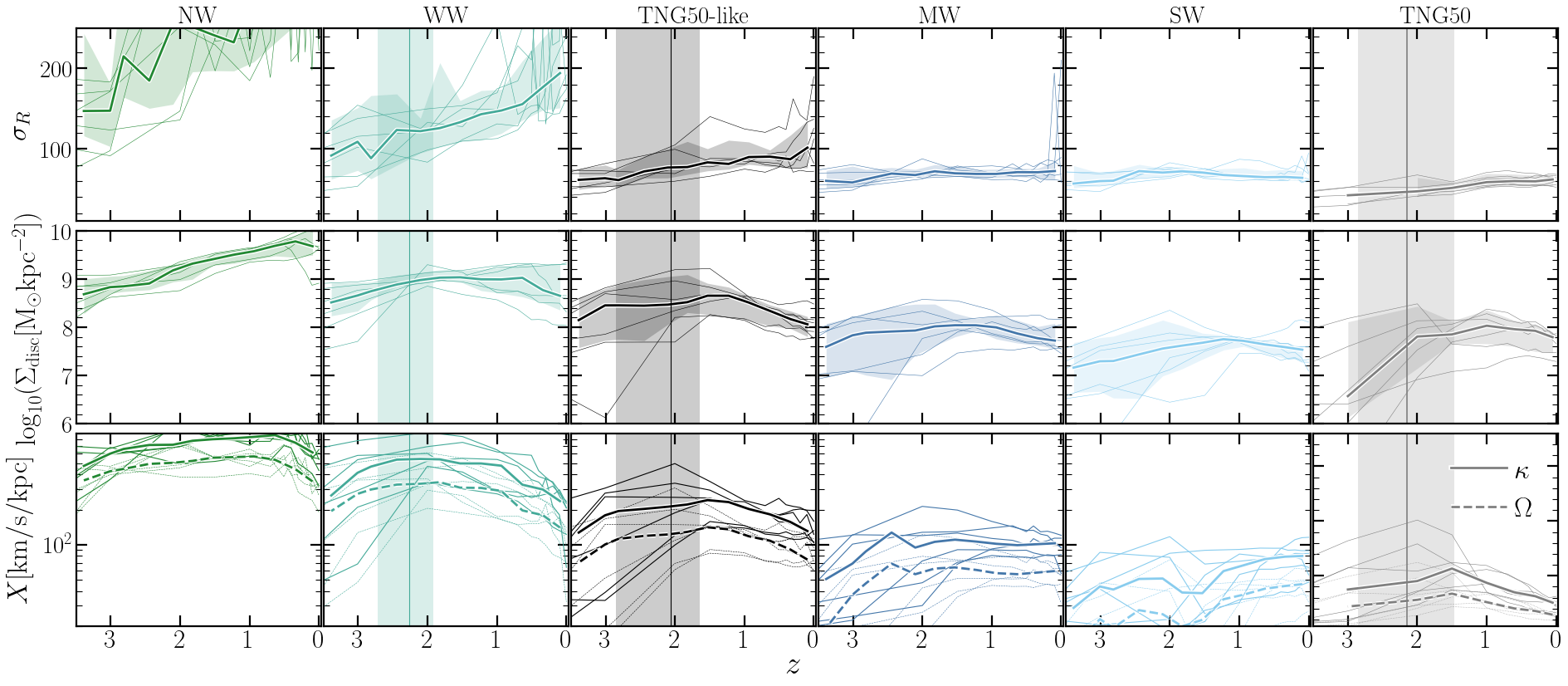}  \\
\includegraphics[width=1.3\columnwidth]{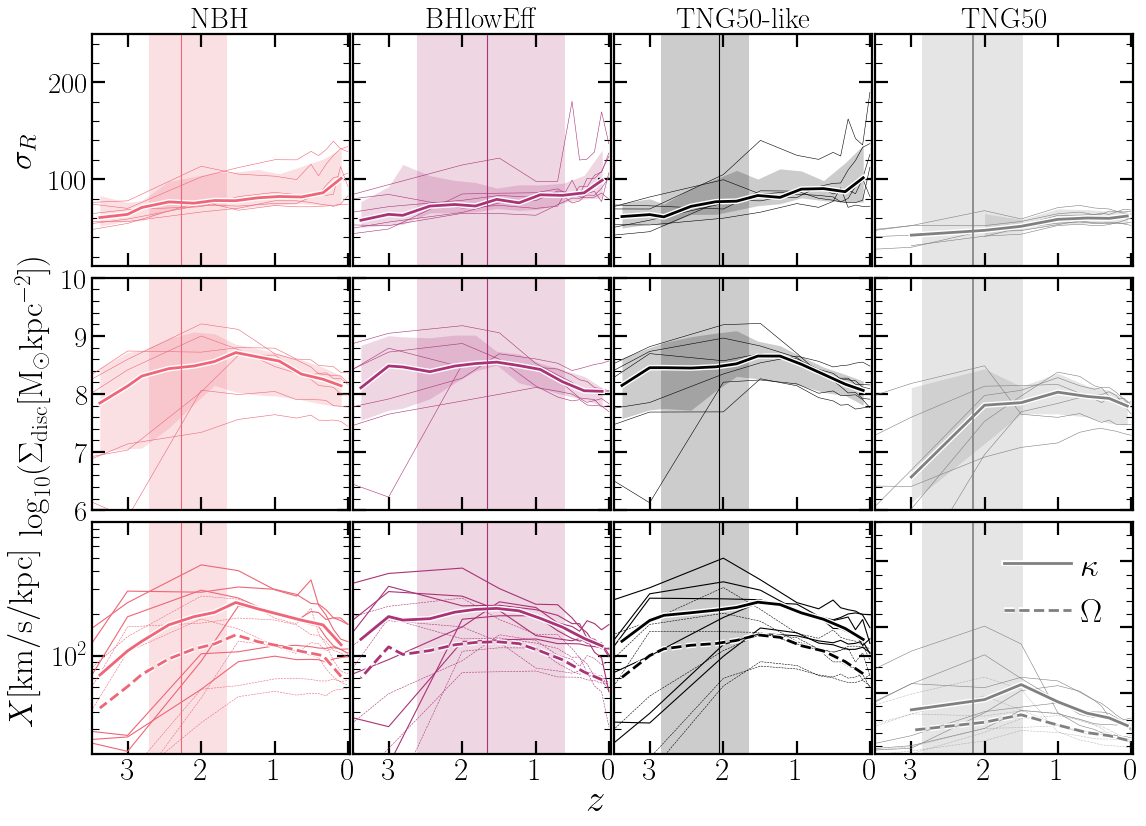}
\end{tabular}
\caption{Evolution of radial velocity dispersion, stellar surface density, epicyclic and angular frequency in a-stellar half-mass radius aperture. From left to right: different galactic wind models (top figure) and BH physics models (bottom figure). The last right Col. corresponds to the original galaxy in TNG50. Thinner solid lines correspond to individual halos. Vertical solid lines and shaded regions represent the median redshift and the distribution between $20^{\rm th}$ and $80^{\rm th}$ percentiles of the bar formation time. The three galaxy properties are highly affected by the SN feedback process, especially the radial velocity dispersion. The highest velocity dispersion is presented in the NW and WW models, causing the Toomre criterion to determine stable discs in contrast to the predictions of ELN and MMW criteria.}
\label{fig:toomre}
\end{figure*}

The bottom panels of Fig.~\ref{fig:ToomreELNLambda} display the average $Q_{\rm T}$ and $X$ for the various models within a stellar half-mass radius and a z-slide of $2$ kpc, which is comparable to the thickness of a disc (refer to Table \ref{table:zooms}). We can observe that the values of $Q_{\rm T}$ and $X$ are located in the bar instability region in the  TNG50-like and BH variation models, particularly during the bar formation phase. It is interesting to note that the values of $Q_{\rm T}$ are just above $2$ at later times. This is in agreement with the values presented in recent observations, which show that some local barred discs can reach values of $Q_{\rm T}$  of $\approx3$ \citep[e.g.][]{romeo2023}.
The swing parameter, $X$, is located in the instability regions of galaxies for all the models, especially in the medium and strong wind models, while $Q_{\rm T}$ is located above the instability region with values larger than $3$ in both the medium and strong wind models (see Cols. 4 and 5 of top figure of Fig. \ref{fig:ToomreELNLambda}). The high values of  $Q_{\rm T}$  determine that the disc is highly stable against bar instabilities. Interestingly, the Toomre parameter is substantially higher in the no/weak wind model for galaxies ($Q_{\rm T}>4$) than in the other models. We will focus on Fig. \ref{fig:toomre} to better understand this behaviour. We note that the galaxies in the no/weak wind models have higher $\kappa$ values. They also present higher stellar surface densities, which could be used to offset the highest values in $\kappa$. Consequently, we obtained similar values in $X$ at $z \lesssim 2$. However, the radial velocity dispersion, which is significantly influenced by the SN feedback, is the primary cause of the discrepancy in $Q_{\rm T}$ between the no/weak wind models and the other models. In addition, this could be the reason for the opposite predictions between this instability criterion, which predicts stable bars for galaxies in NW and WW models, and the ELN and MMW criteria, which predict unstable discs. Even more galaxies in the NW model do not form a stable bar, whereas galaxies in WW galaxies do. The origin of the discrepancy may be because the ELN and MMW criteria do not consider $\sigma_R$  (Eqs. \ref{eq:ELN} \& \ref{eq:MMW}) while $Q_{\rm T}$ does depend on it (Eq.\ref{eq:Toomre}), indicating that the Toomre criterion is sensitive to the precise value of $\sigma_R$. We will discuss this in the next section.

\section{Discussion}
\label{sec:discussion}
\begin{figure*}
\begin{tabular}{c}
\includegraphics[width=2\columnwidth]{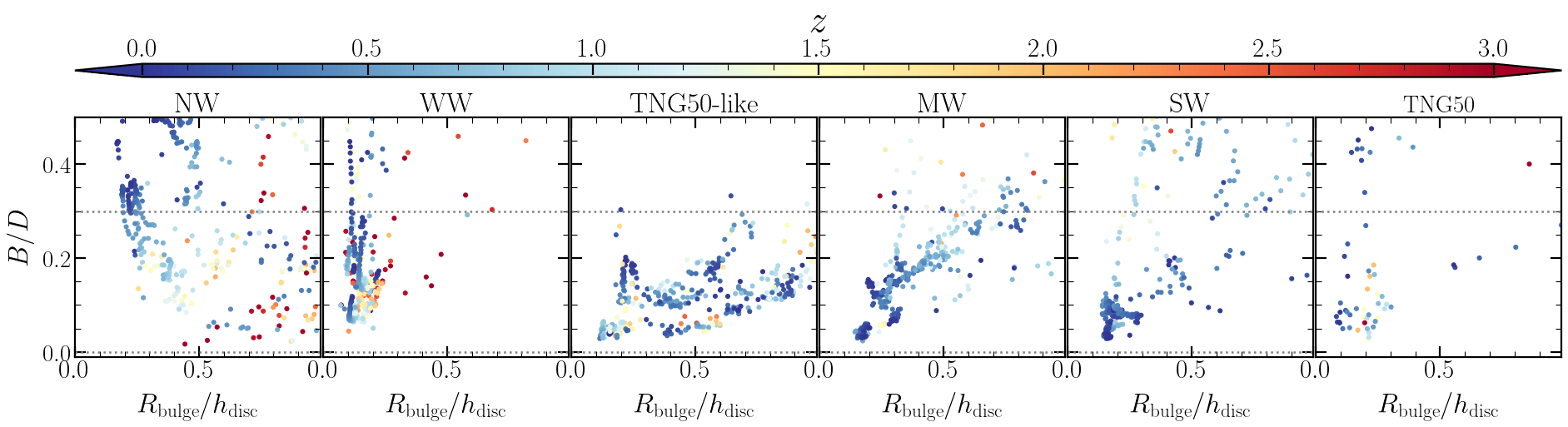}  \\
\includegraphics[width=1.35\columnwidth]{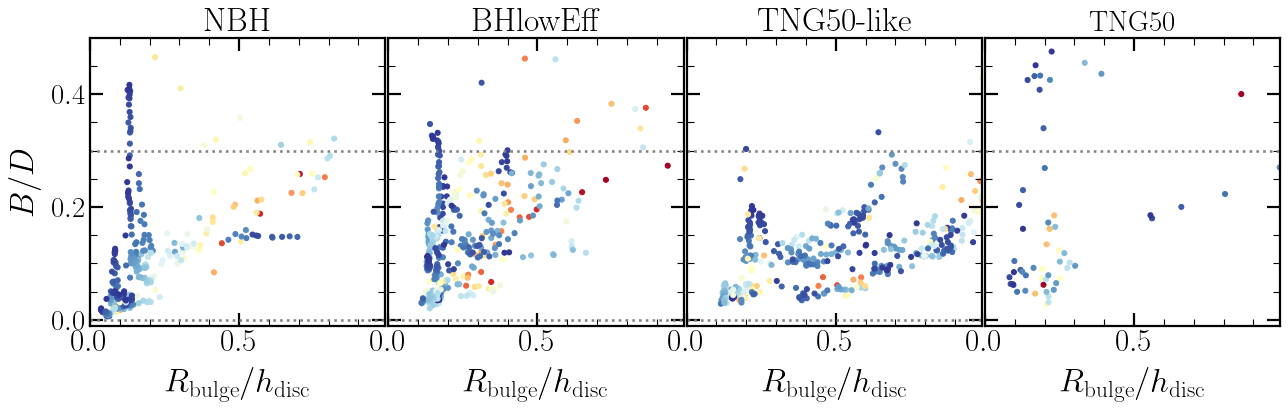} \\
\end{tabular}
\caption{Bulge-to-disc mass fractions as a function of bulge-to-disc size ratios. Left to right: galactic wind models (top Fig) and BH physics model variations (bottom Fig). Colour code represents the redshift. Horizontal grey dashed lines represent the limit given by \cite{kataria2018} for dense bulges that do not favour the formation of a bar, even if the disc could be unstable. Galaxies in the NW model present a large and compact bulge since early times, preventing the disc from forming a bar.}
\label{fig:densitybulge}
\end{figure*}

In this section, we will discuss our findings with respect to previous works and future prospects.

\subsection{Is bar formation affected by the central bulge?}
\label{sub:bulgeeffect}
The ELN and MWW criteria predict that the disc is unstable to bar-like perturbations for galaxies the no wind model; however, a stable bar does not form. These galaxies exhibit the highest stellar mass content in the disc since early times. For example, the findings of \cite{bonoli2016} and \cite{spinoso2017}, which analyse the twin zoom-in simulations of the Milky Way, ErisBH and Eris, suggest that ErisBH forms a bar as a result of AGN feedback, which diminishes the bulge size in comparison to Eris. This enables the activation of bar instabilities. In addition, \cite{saha2018} demonstrate that cold stellar discs surrounded by an initial relative dense bulge ($\rho_{\rm bulge}/\rho_{\rm disc}>1$ for compact bulges) do not form a bar even though the disc was susceptible to bar instabilities (lower values of the Toomre parameter). Similarly, \cite{kataria2018} found that massive and more concentrated bulges can inhibit bar formation and growth, particularly when the bulge-to-disc mass fraction is higher than $0.3$ for concentrated bulges and even higher ($0.6$) for less concentrated bulges. This was achieved using N-body simulations.

Fig. \ref{fig:densitybulge} illustrates the bulge-to-total-mass fraction ($B/D$) as a function of bulge-to-disc sizes ratio and colour-coded by redshift. Galaxies in the no wind model show an increase of $B/D$ with values higher than $0.3$ and bulge-to-disc size ratios smaller than $0.5$ after $z=1$. The upper limit suggested by  \cite{kataria2018}  for compact bulges to prevent bar formation is $B/D=0.3$.  Their simulations indicate that bars can only form in disc galaxies when the radial force due to the bulge is less than $0.35$ in comparison to the total force at the disc scale length. This is an outcome of the bulge's increasing mass, which increases the velocity dispersion of the disc and the radial force. At high redshift ($>1.5$), the galaxies in the no-wind model, generally exhibit $B/D<0.3$; however, the bulge-to-disc size ratios take values higher than $0.5$ (less compact bulges). Consequently, the disc is unstable in principle, but the non-axisymmetric instabilities are unable to grow (see Fig.\ref{fig:evolutionwinds}) since the bulge mass continues to increase constantly (see Fig. \ref{fig:evolutioncomp}).

At the time of bar formation ($z_{\rm bf}\sim2$)  and until  $z\approx 0.5$, the galaxies in the WW model have bulge-to-disc size ratios that are less than $0.5$ and their $B/D$ values are less than $0.3$. This suggests the disc is still capable of forming a bar even though the galaxies in the WW model have concentrated bulges. At low redshift ($z<0.5$), the bulge continues to grow, but the bar length does not increase to the same extent as in the TNG50-like and BH variations models. This may suggest that the bulge could be influencing bar evolution.    In contrast, the medium (MW) and strong (SW) wind models exhibit galaxies that show similar behaviour to those in the TNG50-like model, with $B/D<0.3$ at low redshift. The galaxies present higher $B/D>0.3$, and the bulge-to-disc size ratio takes values higher than $0.5$  at higher redshift. Despite the fact that the bulge may have a minor effect, the discs in the MW and SW models do not form a bar, and according to the three instability criteria, these discs are stable against bar formation in general.

\subsection{ What is the impact of feedback processes in bar formation?}
\label{subsec:feedbackproceses}
As seen in section \ref{sec:galprop}, feedback processes regulate star formation and affect the galaxy build-up and its components.  In particular, stronger SN feedback delays the assembly of the disc (Fig. \ref{fig:evolutioncomp}). As a result,  discs exhibit less mass (Fig. \ref{fig:criterionMo}), lower radial velocity dispersion (Fig. \ref{fig:toomre}) and larger sizes (Fig. \ref{fig:evolutioncomp}), as the supernova feedback strength increases.  These properties are tightly related to determining whether a disc is stable or not (Eqs. \ref{eq:Toomre}, \ref{eq:ELN} and \ref{eq:MMW}). In particular, galaxies in the strong wind model experience more gradual gas cooling and accretion from a hot and diffuse gas halo, than galaxies in the other models \citep[e.g.][]{sales2012}. As a result, galaxies form stable discs. In the case of the galaxies in the No-wind model, insufficient feedback may lead to runaway star formation and rapid consumption of gas, leading to the rapid formation of a massive and compact bulge \citep[e.g.][]{navarro1991}. This has repercussions in the conditions that promote bar formation.

We also note that the galaxies with weak or no-SN feedback are not realistic, and we have only studied them to gain an understanding of the effects of feedback processes on bar formation, but we already know that these models predict galaxies that overestimate the baryon content of galaxies (see \citealt{pillepich2018a}).

Varying quasar BH physics models does not significantly affect the time of disc or the bulge formation (Fig. \ref{fig:evolutioncomp}).  As a result, the mass (Fig. \ref{fig:ELNcriterion}), size (Fig. \ref{fig:evolutioncomp}) and radial dispersion velocity (Fig. \ref{fig:toomre}) are pretty similar. However, the scatter in the time of the formation of the bulge and bar is higher as the efficiency of AGN feedback in quasar mode is lower (Fig. \ref{fig:evolutioncomp}).  This could be because this mode of AGN feedback affects the central gas of the galaxy. Indeed,  \cite{irodotou2022}, which studies a galaxy of the Auriga simulations, has found that bars form with/without quasar mode feedback. This mode of AGN feedback can affect the final bar properties of the bar.

Our findings suggest that the formation of bars may serve as a constraint on the subgrid feedback models of galaxy formation, thereby contributing significantly to our understanding of galaxy evolution. We caution, however, that our galaxy sample, as well as the variations of the subgrid physics, are limited. Therefore, bar formation could be affected by other factors beyond our study, such as galactic surroundings \cite[e.g.][]{peschken2019} which, foster bar formation under given conditions.

\subsection{Lessons learnt from disc instability criteria}
\label{subsec:hintsbf}
As mentioned in section \ref{sec:barformation}, most galaxies among the different models studied here satisfy the three criteria of bar instability. An exception is the galaxies in the NW and WW models. Galaxies in the NW model do not form a bar, while galaxies in the WW model do. Nevertheless, the Toomre criterion predicts stable discs, whereas the ELN and WWM criteria forecast unstable discs. It is possible that a missing condition is required for bar formation, as the ELN and WWM criteria have opposite predictions to the Toomre criteria. Indeed, the galaxies in NW and WW models have formed the most massive and compact bulges of the models studied here, and these components may influence bar formation (Fig. \ref{fig:densitybulge}, section \ref{sub:bulgeeffect}).  The effect of the bulge is not taken into account in the ELN and WWM criteria. This has been studied by \cite{athanassoula2008}  who find that the ELN criterion does not take into account the effect of the central concentration to stabilise the disc. \cite{kataria2018}  using N-body simulations propose a limit on bar formation in discs with bulges by assessing the ratio of radial force due to bulge and disc components at disc scale length. This is because the velocity dispersion of disc stars becomes large enough to prevent bar-type instability.
Similarly, the Toomre parameter also fails to predict a bar in the presence of a massive bulge in galaxies in the weak wind model. The Toomre parameter does not adequately account for the varying effects of different bulge masses (compact vs. less dense) and their interaction with the halo, which can lead to scenarios where bars form or do not form despite being predicted as unstable/stable based on a parameter alone. When a bulge is strong enough, also a disc may not form a bar due to non-linear responses that are not captured by the Toomre criterion. For instance, \cite{jang2023},  using isolated simulation, propose a new condition that includes the central mass concentration (CMC) together with the Toomre parameter to predict if the disc is stable or not.

In addition to this, more ingredients have not been investigated in this study due to the nature of our sample, which comprises galaxies that exhibit bar formation in TNG50 at high redshift ($z>1.5$) and subsequently have low halo spin parameters. For instance, \cite{izquierdo2022} find that a small percentage of barred galaxies in TNG50 do not satisfy the ELN criterion. The potential reason for this is that they had a close encounter with a galaxy during the bar formation process (see also \citealt{rosasguevara2024}). Conversely, other studies have found that close encounters do not promote bar formation, but rather delay bar formation \citep[e.g.][]{zana2018a}.  \cite{algorry2017} also find that the ELN criterion is not sufficient to predict stable bars in the EAGLE simulation and proposed an additional parameter which evaluates the dynamical contribution of the baryons at a half mass radius scale with respect to the global dynamics of the dark matter halo (maximum circular velocity of the halo).

In conclusion, there is no consensus on the importance and the limits of secondary conditions needed to promote/prevent bar formation, maybe reflecting the complex processes that are involved in bar formation.

\subsection{What is the effect of stochasticity of feedback processes in bar formation}

Something to be aware of when trying to understand our results is to take into account the stochasticity provided by subgrid physics in simulations. In particular, \cite{genel2019} found that simulations with feedback mechanisms could exhibit a stronger butterfly effect than those without feedback. This suggests that the dynamics at galactic scales could significantly influence how small differences in initial conditions can evolve into larger discrepancies in galaxy properties (see also \citealt{borrow2023} for other galaxy formation models). It is essential to acknowledge that the morphology of the models under investigation here is not significantly affected; however, the radial velocity dispersion and the extent of the disc could be. These qualities are employed to determine the stability of a disc, as seen in Figs. \ref{fig:evolutioncomp} \& \ref{fig:toomre}. The TNG50 simulation predicted the presence of six strongly barred galaxies; however, only five of these galaxies present a strong bar in the TNG50-like model at $z=0$. For this particular galaxy, we note that the original TNG50 galaxy has a strong bar with the weakest strength among the six galaxies studied here ($A_{2,\rm max}~0.38$, see Table~\ref{table:haloes}) at $z=0$. In the TNG50-like model, the galaxy forms a bar that appears to weaken over time, as it evolves. We suspect that the cause of this weakening is the fact that the galaxy in the TNG50-like model has higher gas fractions in the centre than the original TNG50 galaxy. This could be because the black hole is less massive, and the energy released in radio mode is less. This is an example of how the stochasticity of subgrid physics could affect bar formation, but in general, we find minor variations in the characteristics of strong bars in our study.
On the other hand, when the feedback mechanisms differ, the merging histories are very similar to the original TNG50. The stochastic uncertainty that is incorporated into simulations limits the physical conclusions that can be derived. The broad causal tendencies we have identified suggest that this constraint is of lesser significance. Finally, we would like to emphasise that our study is limited to five galaxies with a strong bar, which does not allow the drawing of general conclusions regarding the effects of the stochasticity of feedback processes and a more comprehensive study in this matter using larger cosmological simulations with a larger galaxy statistical sample is needed for future projects, especially for the weak bars.

\subsection{Outlook}
The zoom-in hydrodynamic simulations have enabled us to investigate the influence of feedback processes with the advantage of previous works using the TNG50 simulation. We resimulate six Milky-Way like galaxies from the TNG50 simulations. These galaxies have a stable galactic bar at $z=0$  which was formed between $z=3$ and $z=1.5$  (more than 8 gigayears ago) and with quiet merger histories and isolated haloes.  Even so, it is crucial to recognise the limitations of this investigation. The physics above scales of approximately $288$ pc can be resolved by these zoom-ins. Although this is sufficient for a comprehensive examination of the formation and evolution of bars within the most massive galaxies in the local universe, simulations with a higher resolution are required to correctly track the initial phases of bar build-up at high redshifts. Recent observations from JWST have shown that bars exist at high redshifts \citep[e.g.][]{guo2023,costantin2023,guo2024}.
The conditions which galaxies are subject to at such high redshifts are notably distinct from lower redshifts.  It is anticipated that the frequency of encounters and mergers is higher  at such early times \citep[e.g.][]{conselice2008,bi2022a,bi2022b}. A cosmological context would be an interesting setting in which to conduct this experiment with an extensive statistical galaxy sample and in different environments at higher redshifts.

It is thought that bars drive flows into the galaxy centre, which in turn affects central star formation or the accretion onto the supermassive black hole  (e.g. \citealt{fanali2015,spinoso2017}). The fate of the gas in the nuclear zone, affected by the bar and other physical processes like SN and AGN feedback, should be studied. In particular, it would be interesting to see gaseous bars formed by stellar bar torques \citep[e.g.][]{englmaier2004}.

The effects of the star formation processes on the bar properties, especially the bar pattern speed, which is the least constrained, need to be studied. The bar pattern speed is most affected by the interaction between the bar, disc, and the central dark matter distribution \citep[e.g.][]{cuomo2020}. The main effect is a slowdown of the bar as a function of time. This seems to be in contradiction with observations, which have found that bar pattern speeds are rapid. Using the zoom-in simulations, we can see the effect of the star formation processes at high redshift that affect the initial bar pattern speed when the bar forms \citep[e.g.][]{semczuk2024}.


\section{Summary}
\label{sec:summary}

In this work, we investigate the impact of feedback processes on the formation and evolution of bars in galaxies similar to the Milky Way. We study a suite of zoom-in cosmological hydrodynamical simulations of Milky Way-sized haloes selected from the TNG50 simulations in which the original galaxies form a strong bar between $z=1.5$ and $z=3$ 
In particular, we explore variations of the supernova (SN) feedback and black hole (BH) physics models within the TNG galaxy formation model. Specifically, our investigation focuses on five different intensity levels within the galactic wind models (energy injected into the galactic wind per SN event) and two models of black hole physics affecting the AGN feedback in the quasar mode regime.

The results of our investigation are as follows:
\begin{itemize}
\item Our analysis reveals that the feedback from SN impacts the formation of a bar, while the effect of the model variations of quasar mode BH physics studied here is negligible (Figs. \ref{fig:DensityMap} \& \ref{fig:evolutionwinds}). However, BH physics can affect the ultimate characteristics of a bar, such as strength and length. In order to gain a deeper understanding, we examine the characteristics of the various components of the galaxy.
 \item  It is observed that the morphology at low redshift remained consistent in the sense that in all the models, galaxies developed a massive disc ($>10^{10}\Msun$),  which is dominant in comparison with the bulge mass (Fig. \ref{fig:evolutionwinds}).
\item Our findings indicate that SN feedback has a significant impact not only on the stellar mass content but also on the structural properties of the disc and bulge, including the disc size, thickness, and bulge size (see Table \ref{table:zooms} \& Fig.\ref{fig:evolutioncomp}). SN feedback also affects the formation time of the disc and bulge, and then bar formation, which aligns with the finding by \cite{zana2019}. In extreme cases where the SN feedback is strong, bar formation is stopped. Our analysis revealed that the bulge and disc in galaxies lacking SN feedback are noticeably more compact and concentrated than those with stronger feedback at $z=0$. In such galaxies, bar formation also stops.

\item Based on these findings, it is evident that the feedback from supernovae plays a crucial role in shaping the various components found within galaxies. When the feedback strength is increased, it causes a decrease in the rate at which the disc grows. As a result, this affects the (thin) disc and bulge characteristics. The models varying BH physics in quasar mode do not influence the disc or the bulge formation, as also seen in \cite{irodotou2022} using the Auriga simulations.

\item We thoroughly review the bar instability criteria in our different models,  in particular during the process of bar formation. Specifically,  we focus on three criteria: the \citealt{toomre1964}, \citet[][ELN]{ efstathiou1982}, and \citet[][MMW]{mo1998} criteria. Our analysis revealed that the sizes of the discs (Fig.~\ref{fig:evolutioncomp}), the stellar mass of the disc over halo mass (Fig.~\ref{fig:criterionMo}) and the radial velocity dispersion (Fig.~\ref{fig:toomre}) are strongly influenced by SN feedback and used to determine if a disc is prone to instability and bar formation (Fig.\ref{fig:ToomreELNLambda}). Most of the models satisfy the instability criteria at the moment of bar formation, as in the barred galaxies in  TNG50 \citep{izquierdo2022}. The exception is the no wind model, which presents higher radial velocity dispersion, leading to high values of the Toomre parameter, whereas the ELT and MMW criteria predict an unstable disc.

\item In galaxies in the weak winds and TNG50-like models, discs exhibit instability and give rise to bar formation, while galaxies in the strong and medium wind models do not form a bar. On the other hand, galaxies in the no wind model do not form a bar even though the disc satisfies some instability criteria. This could be explained by the presence of a massive and compact bulge (see Fig.~\ref{fig:densitybulge}). This is in agreement with \cite{kataria2018}, which found that galaxies with $B/D>0.3$  for concentrated bulges ($B/D>0.6$ for less concentrated bulges) prevent bar formation.

\end{itemize}

 The comparison of galaxy formation models in our study offers a deeper understanding of the relationship between the emergent structures of galaxies and the input physics assumptions during their formation and evolution. This understanding could inspire new approaches to the prescription of galaxy formation models, offering hope for further advancements in the field.

Our research can also be employed to evaluate the predictive capabilities of analytics methods employed to identify the presence of a bar. This would enhance our comprehension of bar formation and enhance the modelling of barred galaxies in more empirical models.

\begin{acknowledgements}
The authors thank Volker Springel for providing them with the \textsc{Arepo} code and the TNG model and his suggestions regarding the project. The authors would like to thank the anonymous referee for their comments which improved the quality of this paper. The authors thankfully acknowledge the computer resources at MareNostrum and the technical support provided by Barcelona Supercomputing Centre (RES-AECT-2023-2-0015). YRG acknowledges the support of the``Juan de la Cierva Incorporacion'' Fellowship (IJC2019-041131-I). S.B. \& YRG  acknowledge support from the Spanish Ministerio de Ciencia e Innovación through project PID2021-124243NB-C21.  SC acknowledges the support of the `Juan de la Cierva Incorporac\'ion' fellowship (IJC2020-045705-I). S.B., SC \& YRG acknowledge support from the European Research Executive Agency HORIZON-MSCA-2021-SE-01 Research and Innovation Programme under the Marie Sklodowska-Curie grant agreement number 101086388 (LACEGAL). Technical and human support provided by the DIPC Supercomputing Centre is gratefully acknowledged.
\end{acknowledgements}

%

\begin{thebibliography}{99}



\bibitem[\protect\citeauthoryear{Algorry et al.}{2017}]{algorry2017}
Algorry D.~G., et al., 2017, MNRAS, 469, 1054

\bibitem[\protect\citeauthoryear{Aric{\`o} et al.}{2021}]{arico2021} Aric{\`o} G., Angulo R.~E., Hern{\'a}ndez-Monteagudo C., Contreras S., Zennaro M., 2021, MNRAS, 503, 3596.



\bibitem[\protect\citeauthoryear{Athanassoula \& Misiriotis}{2002}]{athanassoula2002}
Athanassoula E., Misiriotis A., 2002, MNRAS, 330, 35

\bibitem[\protect\citeauthoryear{Athanassoula}{2003}]{athanassoula2003}
Athanassoula E. 2003, MNRAS, 341, 1179 (A03)


\bibitem[\protect\citeauthoryear{Athanassoula }{2008}]{athanassoula2008}
Athanassoula E., 2008, MNRAS, 390, L69.


\bibitem[\protect\citeauthoryear{Athanassoula et al.}{2013}]{athanassoula2013}
Athanassoula E., Machado R.~E.~G., Rodionov S.~A., 2013, MNRAS, 429, 1949



\bibitem[\protect\citeauthoryear{Bi, Shlosman, \& Romano-D{\'\i}az}{2022a}]{bi2022a} Bi D., Shlosman I., Romano-D{\'\i}az E., 2022, MNRAS, 513, 693.

\bibitem[\protect\citeauthoryear{Bi, Shlosman, \& Romano-D{\'\i}az}{2022b}]{bi2022b} Bi D., Shlosman I., Romano-D{\'\i}az E., 2022, ApJ, 934, 52.

\bibitem[\protect\citeauthoryear{Binney \& Tremaine}{2008}]{binney2008}
 Binney J., Tremaine S., 2008, gady.book

\bibitem[\protect\citeauthoryear{Bird et al.}{2014}]{bird2014} Bird S., Vogelsberger M., Haehnelt M., Sijacki D., Genel S., Torrey P., Springel V., et al., 2014, MNRAS, 445, 2313.


\bibitem[\protect\citeauthoryear{Bonoli et al.}{2016}]{bonoli2016}
Bonoli S., Mayer L., Kazantzidis S., Madau P., Bellovary J., Governato F., 2016, MNRAS, 459, 2603

\bibitem[\protect\astroncite{Bower et al.}{2006}]{bower2006} Bower R.~G.,
Benson A.~J., Malbon R., Helly J.~C., Frenk C.~S., Baugh C.~M., Cole S., Lacey
C.~G., 2006, MNRAS, 370, 645.

\bibitem[\protect\citeauthoryear{Borrow et al.}{2023}]{borrow2023}
Borrow J., Schaller M., Bah{\'e} Y.~M., Schaye J., Ludlow A.~D., Ploeckinger S., Nobels F.~S.~J., et al., 2023, MNRAS, 526, 2441.

\bibitem[\protect\citeauthoryear{Bullock et al.}{2001}]{bullock2001}
Bullock J.~S., Dekel A., Kolatt T.~S., Kravtsov A.~V., Klypin A.~A., Porciani C., Primack J.~R., 2001, ApJ, 555, 240.



\bibitem[\protect\citeauthoryear{Cervantes-Sodi et al.}{2015}]{cervantes2015}
Cervantes Sodi B., Li C., Park C., 2015, ApJ, 807, 111


\bibitem[\protect\citeauthoryear{Cervantes-Sodi}{2017}]{cervantes2017}
Cervantes Sodi B., 2017, ApJ, 835, 80

\bibitem[\protect\citeauthoryear{Conselice, Rajgor, \& Myers}{2008}]{conselice2008}
Conselice C.~J., Rajgor S., Myers R., 2008, MNRAS, 386, 909.


\bibitem[\protect\citeauthoryear{Collier, Shlosman, \& Heller}{2018}]{collier2018} Collier A., Shlosman I., Heller C., 2018, MNRAS, 476, 1331.




\bibitem[\protect\citeauthoryear{Costantin et al.}{2023}]{costantin2023}
Costantin L., P{\'e}rez-Gonz{\'a}lez P.~G., Guo Y., Buttitta C., Jogee S., Bagley M.~B., Barro G., et al., 2023, Natur, 623, 499

\bibitem[\protect\citeauthoryear{Crain et al.}{2023}]{crain2023}
Crain R.~A., van de Voort F., 2023, ARA\&A, 61, 473.

\bibitem[\protect\astroncite{Croton et al.}{2006}]{croton2006}
Croton D. ~J., Springel V., White S. ~D. ~M., De Lucia ~G., Frenk S. C., Gao L., Jenkins ~A. and Kauffmann ~G. {\it et al.}, 2006, MNRAS, 365, 11.

\bibitem[\protect\citeauthoryear{Cuomo et al.}{2020}]{cuomo2020} Cuomo V., Aguerri J.~A.~L., Corsini E.~M., Debattista V.~P., 2020, A\&A, 641, A111.
\bibitem[\protect\citeauthoryear{Dahari}{1984}]{dahari1984}
Dahari O., 1984, The Astronomical Journal, 89, 966


\bibitem[\protect\citeauthoryear{Jang \& Kim}{2023}]{jang2023} Jang D., Kim W.-T., 2023, ApJ, 942, 106.




\bibitem[\protect\citeauthoryear{Debattista et al.}{2006}]{debattista2006} Debattista V.~P., Mayer L., Carollo C.~M., Moore B., Wadsley J., Quinn T., 2006, ApJ, 645, 209





\bibitem[\protect\citeauthoryear{Di Matteo et al.}{2005}]{diMatteo2005} Di Matteo T., Springel V., Hernquist L., 2005, Natur, 433, 604

\bibitem[\protect\citeauthoryear{Donohoe-Keyes, et al.}{2019}]{donohoe2019} Donohoe-Keyes C.~E., Martig M., James P.~A., Kraljic K., 2019, arXiv, arXiv:1908.11119



\bibitem[\protect\citeauthoryear{Efstathiou, Lake \& Negroponte}{1982}]{efstathiou1982}
Efstathiou G., Lake G., Negroponte J., 1982, MNRAS, 199, 1069


\bibitem[\protect\citeauthoryear{Englmaier \& Shlosman}{2004}]{englmaier2004} Englmaier P., Shlosman I., 2004, ApJL, 617, L115.






\bibitem[\protect\citeauthoryear{Fanali et al.}{2015}]{fanali2015}
Fanali R., Dotti M., Fiacconi D., Haardt F., 2015, MNRAS, 454, 3641



\bibitem[\protect\citeauthoryear{Fragkoudi et al.}{2020}]{fragkoudi2020}
Fragkoudi F., Grand R.~J.~J., Pakmor R., Bl{\'a}zquez-Calero G., Gargiulo I., Gomez F., Marinacci F., et al., 2020, MNRAS, 494, 5936.

\bibitem[\protect\citeauthoryear{Fragkoudi et al.}{2024}]{fragkoudi2024} Fragkoudi F., Grand R., Pakmor R., G{\'o}mez F., Marinacci F., Springel V., 2024, arXiv, arXiv:2406.09453.

\bibitem[\protect\citeauthoryear{Fraser-McKelvie et al.}{2020}]{fraser2020} Fraser-McKelvie A., Merrifield M., Arag{\'o}n-Salamanca A., Peterken T., Kraljic K., Masters K., Stark D., et al., 2020, MNRAS, 499, 1116.


\bibitem[\protect\citeauthoryear{Faucher-Gigu{\`e}re, et al.}{2009}]{faucher2009} Faucher-Gigu{\`e}re C.-A., Lidz A., Zaldarriaga M., Hernquist L., 2009, ApJ, 703, 1416


\bibitem[\protect\citeauthoryear{Gadotti}{2009}]{gadotti2009}
Gadotti D.~A., 2009, MNRAS, 393, 1531






\bibitem[\protect\citeauthoryear{Gavazzi et al.}{2015}]{gavazzi2015}
Gavazzi G., et al., 2015, A\&A, 580, A116




\bibitem[\protect\citeauthoryear{{Genel} et~al.}{{Genel}   et~al.}{2014}]{genel2014}
Genel S., et al., 2014, MNRAS, 445, 175



\bibitem[\protect\citeauthoryear{Genel et al.}{2019}]{genel2019}
Genel S., Bryan G.~L., Springel V., Hernquist L., Nelson D., Pillepich A., Weinberger R., et al., 2019, ApJ, 871, 21

\bibitem[\protect\citeauthoryear{George et al.}{2019}]{george2019}
George K., Subramanian S., Paul K.~T., 2019, arXiv, arXiv:1907.06910



\bibitem[\protect\citeauthoryear{Ghosh et al.}{2023}]{ghosh2023}
Ghosh S., Fragkoudi F., Di Matteo P., Saha K., 2023, A\&A, 674, A128.

\bibitem[\protect\citeauthoryear{Guedes et al.}{2011}]{guedes2011}
Guedes J., Callegari S., Madau P., Mayer L., 2011, ApJ, 742, 76

\bibitem[\protect\citeauthoryear{Guo et al.}{2023}]{guo2023}
Guo Y., Jogee S., Finkelstein S.~L., Chen Z., Wise E., Bagley M.~B., Barro G., et al., 2023, ApJL, 945, L10.
\bibitem[\protect\citeauthoryear{Guo et al.}{2024}]{guo2024} Guo Y., Jogee S., Wise E., Pritchett K., McGrath E.~J., Finkelstein S.~L., Iyer K.~G., et al., 2024,
\bibitem[\protect\citeauthoryear{Grand et al.}{2017}]{grand2017}
Grand R.~J.~J., G{\'o}mez F.~A., Marinacci F., Pakmor R., Springel V., Campbell D.~J.~R., Frenk C.~S., et al., 2017, MNRAS, 467, 179.


\bibitem[\protect\citeauthoryear{Henriques et al.}{2013}]{henriques2013} Henriques B.~M.~B., White S.~D.~M., Thomas P.~A., Angulo R.~E., Guo Q., Lemson G., Springel V., 2013, MNRAS, 431, 3373.


\bibitem[\protect\citeauthoryear{Irodotou et al.}{2022}]{irodotou2022} Irodotou D., Fragkoudi F., Pakmor R., Grand R.~J.~J., Gadotti D.~A., Costa T., Springel V., et al., 2022, MNRAS, 513, 3768.



\bibitem[\protect\citeauthoryear{Izquierdo-Villalba et al.}{2022}]{izquierdo2022}
Izquierdo-Villalba D., Bonoli S., Rosas-Guevara Y., Springel V., White S.~D.~M., Zana T., Dotti M., et al., 2022, arXiv, arXiv:2203.07734

\bibitem[\protect\citeauthoryear{Joshi et al}{2024}]{joshi2024}
Joshi G.~D., Pontzen A., Agertz O., Rey M.~P., Read J., Pillepich A., 2024, arXiv, arXiv:2407.00171.





\bibitem[\protect\citeauthoryear{Kataria \& Das}{2018}]{kataria2018}
Kataria S.~K., Das M., 2018, MNRAS, 475, 1653

\bibitem[\protect\citeauthoryear{Katz et al.}{1996}]{katz1996}
Katz N., Weinberg D.~H., Hernquist L., 1996, ApJS, 105, 19


\bibitem[\protect\citeauthoryear{Khoperskov et al.}{2024}]{khoperskov2024} Khoperskov S., Minchev I., Steinmetz M., Ratcliffe B., Walcher J.~C., Libeskind N.~I., 2024, MNRAS, 533, 3975.



\bibitem[\protect\citeauthoryear{Le Conte et al.}{2023}]{leconte2023}
Le Conte Z.~A., Gadotti D.~A., Ferreira L., Conselice C.~J., de S{\'a}-Freitas C., Kim T., Neumann J., et al., 2023, arXiv, arXiv:2309.10038


\bibitem[\protect\citeauthoryear{Kraljic et al.}{2012}]{kraljic2012}
Kraljic K., Bournaud F., Martig M., 2012, ApJ, 757, 60






\bibitem[\protect\citeauthoryear{Long, Shlosman, \& Heller}{2014}]{long2014}
Long S., Shlosman I., Heller C., 2014, ApJL, 783, L18










\bibitem[\protect\citeauthoryear{Marinacci et al.}{2018}]{marinacci2018}
Marinacci F., et al., 2018, MNRAS, 480, 5113






\bibitem[\protect\citeauthoryear{Masters et al.}{2011}]{masters2011}
Masters K.~L., et al., 2011, MNRAS, 411, 2026







\bibitem[\protect\citeauthoryear{Mo, Mao, \& White}{1998}]{mo1998} Mo H.~J., Mao S., White S.~D.~M., 1998, MNRAS, 295, 319.


\bibitem[\protect\citeauthoryear{Naiman et~al.}{2018}]{naiman2018}
Naiman J. P., et al., 2018, MNRAS, 477, 1206

\bibitem[\protect\citeauthoryear{Navarro \& Benz}{1991}]{navarro1991} Navarro J.~F., Benz W., 1991, ApJ, 380, 320.


\bibitem[\protect\citeauthoryear{Nelson et al.}{2015}]{nelson2015}
Nelson D., et al., 2015, A\& C, 13, 12

\bibitem[\protect\citeauthoryear{Nelson et~al.}{2018}]{nelson2018}
Nelson D., et al., 2018a, MNRAS, 475, 624


\bibitem[\protect\citeauthoryear{Nelson, et al.}{2019a}]{nelson2019a}
Nelson D., et al., 2019a, ComAC, 6, 2

\bibitem[\protect\citeauthoryear{Nelson et al.}{2019b}]{nelson2019b}
Nelson D., Pillepich A., Springel V., Pakmor R., Weinberger R., Genel S., Torrey P., et al., 2019b, MNRAS, 490, 3234.


\bibitem[\protect\citeauthoryear{Oppenheimer \& Dav{\'e}}{2006}]{oppenheimer2006} Oppenheimer B.~D., Dav{\'e} R., 2006, MNRAS, 373, 1265.


\bibitem[\protect\citeauthoryear{Oppenheimer \& Dav{\'e}}{2008}]{oppenheimer2008} Oppenheimer B.~D., Dav{\'e} R., 2008, MNRAS, 387, 577.

\bibitem[\protect\citeauthoryear{Ostriker \& Peebles}{1973}]{ostriker1973}
Ostriker J.~P., Peebles P.~J.~E., 1973, ApJ, 186, 467


\bibitem[\protect\citeauthoryear{Peschken \& Lokas}{2019}]{peschken2019}
Peschken N., {\L}okas E.~L., 2019, MNRAS, 483, 2721

\bibitem[\protect\citeauthoryear{Pillepich et~al.,}{2018a}]{pillepich2018a}
Pillepich A., et al., 2018a, MNRAS, 473, 4077

\bibitem[\protect\citeauthoryear{Pillepich et~al.,}{2018b}]{pillepich2018b}
Pillepich A., et al., 2018b, MNRAS, 475, 648

\bibitem[\protect\citeauthoryear{Pillepich et al.}{2019}]{pillepich2019}
Pillepich A., Nelson D., Springel V., Pakmor R., Torrey P., Weinberger R., Vogelsberger M., et al., 2019, MNRAS, 490, 3196

\bibitem[\protect\citeauthoryear{Pillepich et al.}{2023}]{pillepich2023} Pillepich A., Sotillo-Ramos D., Ramesh R., Nelson D., Engler C., Rodriguez-Gomez V., Fournier M., et al., 2023, arXiv, arXiv:2303.16217.

\bibitem[\protect\citeauthoryear{{Planck Collaboration}}{{Planck Collaboration} }{2016}]{planck2016}
Planck Collaboration 2016, A\& A, 594, A13

\bibitem[\protect\citeauthoryear{Pontzen \& Governato}{2012}]{pontzen2012}
Pontzen A., Governato F., 2012, MNRAS, 421, 3464.






\bibitem[\protect\citeauthoryear{{Rodriguez-Gomez} et~al.,}{{Rodriguez-Gomez} et~al.}{2015}]{rodriguezgomez2015}
Rodriguez-Gomez V., et al., 2015, MNRAS, 449, 49


\bibitem[\protect\citeauthoryear{Romeo, Agertz, \& Renaud}{2023}]{romeo2023} Romeo A.~B., Agertz O., Renaud F., 2023, MNRAS, 518, 1002.
\bibitem[\protect\citeauthoryear{Rosito et al.}{2019}]{rosito2019} Rosito M.~S., Tissera P.~B., Pedrosa S.~E., Rosas-Guevara Y., 2019, A\&A, 629, A37.

\bibitem[\protect\citeauthoryear{Rosas-Guevara et al.}{2020}]{rosasguevara2020}
Rosas-Guevara Y., Bonoli S., Dotti M., Zana T., Nelson D., Pillepich A., Ho L.~C., et al., 2020, MNRAS, 491, 2547
\bibitem[\protect\citeauthoryear{Rosas-Guevara et al.}{2022}]{rosasguevara2022}
Rosas-Guevara Y., Bonoli S., Dotti M., Izquierdo-Villalba D., Lupi A., Zana T., Bonetti M., et al., 2022, MNRAS, 512, 5339

\bibitem[\protect\citeauthoryear{Rosas-Guevara et al.}{2024}]{rosasguevara2024} Rosas-Guevara Y., Bonoli S., Misa Moreira C., Izquierdo-Villalba D., 2024, A\&A, 684, A179.
\bibitem[\protect\citeauthoryear{Saha \& Naab}{2013}]{saha2013}
Saha K., Naab T., 2013, MNRAS, 434, 1287.

\bibitem[\protect\citeauthoryear{Saha \& Elmegreen}{2018}]{saha2018} Saha K., Elmegreen B., 2018, ApJ, 858, 24.

\bibitem[\protect\citeauthoryear{Sales et al.}{2012}]{sales2012}
Sales L.~V., Navarro J.~F., Theuns T., Schaye J., White S.~D.~M., Frenk C.~S., Crain R.~A., et al., 2012, MNRAS, 423, 1544.
\bibitem[\protect\citeauthoryear{Scannapieco \& Athanassoula}{2012}]{scannapieco2012}
Scannapieco C., Athanassoula E., 2012, MNRAS, 425, L10

\bibitem[\protect\citeauthoryear{Schaye et al.}{2015}]{schaye2015}
Schaye J., et al., 2015, MNRAS, 446, 521

\bibitem[\protect\citeauthoryear{Sellwood \& Wilkinson}{1993}]{sellwood1993}
Sellwood J.~A., Wilkinson A., 1993, RPPh, 56, 173


\bibitem[\protect\citeauthoryear{Sellwood}{2012}]{sellwood2012}
Sellwood J.~A., 2012, ApJ, 751, 44.





\bibitem[\protect\citeauthoryear{Sheth et al.}{2012}]{sheth2012}
Sheth K., Melbourne J., Elmegreen D.~M., Elmegreen B.~G., Athanassoula E., Abraham R.~G., Weiner B.~J., 2012, ApJ, 758, 136


\bibitem[\protect\citeauthoryear{Sijacki et al.}{2015}]{sijacki2015}
Sijacki D., et al., 2015, MNRAS, 452, 575

\bibitem[\protect\citeauthoryear{Semczuk et al.}{2024}]{semczuk2024} Semczuk M., Dehnen W., Sch{\"o}nrich R., Athanassoula E., 2024, arXiv, arXiv:2407.11154.

\bibitem[\protect\citeauthoryear{Spinoso et al.}{2017}]{spinoso2017}
Spinoso D., Bonoli S., Dotti M., Mayer L., Madau P., Bellovary J., 2017, MNRAS, 465, 3729


\bibitem[\protect\citeauthoryear{{Springel}}{{Springel}}{2001}]{springel2001}
Springel V., White S. D. M., Tormen G., Kauffmann G., 2001, MNRAS, 328, 726

\bibitem[\protect\citeauthoryear{Springel \& Hernquist}{2003}]{springel2003}
Springel V., Hernquist L., 2003, MNRAS, 339, 289

\bibitem[\protect\citeauthoryear{Springel et al.}{2005}]{springel2005}
Springel V., Di Matteo T., Hernquist L., 2005, MNRAS, 361, 776


\bibitem[\protect\citeauthoryear{{Springel}}{{Springel}}{2010}]{springel2010}
Springel V., 2010, MNRAS, 401, 791



\bibitem[\protect\citeauthoryear{Springel}{2015}]{2015ascl.soft02003S}
Springel V., 2015, ascl.soft. ascl:1502.003


\bibitem[\protect\citeauthoryear{Springel et al.}{2018}]{springel2018}
Springel V., et al., 2018, MNRAS, 475, 676


\bibitem[\protect\citeauthoryear{Suresh et al.}{2015}]{suresh2015}
Suresh J., Bird S., Vogelsberger M., Genel S., Torrey P., Sijacki D., Springel V., et al., 2015, MNRAS, 448, 895.

\bibitem[\protect\citeauthoryear{Toomre}{1964}]{toomre1964} Toomre A., 1964, ApJ, 139, 1217.




\bibitem[\protect\citeauthoryear{Vogelsberger, et al.}{2013}]{vogelsberger2013}
Vogelsberger M., Genel S., Sijacki D., Torrey P., Springel V., Hernquist L., 2013, MNRAS, 436, 3031


\bibitem[\protect\citeauthoryear{Vogelsberger et~al.}{2014a}]{vogelsberger2014a}
Vogelsberger M., et al., 2014a, Nature, 509, 177

\bibitem[\protect\citeauthoryear{Vogelsberger et al.}{2014b}]{vogelsberger2014b}
Vogelsberger M., et al., 2014b, MNRAS, 444, 1518

\bibitem[\protect\citeauthoryear{Vogelsberger et al.}{2020a}]{vogelsberger2020a}
Vogelsberger M., Marinacci F., Torrey P., Puchwein E., 2020, NatRP, 2, 42.


\bibitem[\protect\citeauthoryear{Weinberger et~al.}{{Weinberger} et~al.}{2017}]{weinberger2017}
Weinberger R., et al., 2017, MNRAS, 465, 3291



\bibitem[\protect\citeauthoryear{Yurin \& Springel}{2015}]{yurin2015}
Yurin D., Springel V., 2015, MNRAS, 452, 2367.

\bibitem[\protect\citeauthoryear{Zana et al.}{2018a}]{zana2018a}
Zana T., Dotti M., Capelo P.~R., Bonoli S., Haardt F., Mayer L., Spinoso D., 2018a, MNRAS, 473, 2608




\bibitem[\protect\citeauthoryear{Zana et al.}{2019}]{zana2019}
Zana T., et al., 2019, MNRAS, 488, 1864

\bibitem[\protect\citeauthoryear{Zana et al.}{2022}]{zana2022}
Zana T., Lupi A., Bonetti M., Dotti M., Rosas-Guevara Y., Izquierdo-Villalba D., Bonoli S., et al., 2022, MNRAS, 515, 1524.

\bibitem[\protect\citeauthoryear{Zhao et al.}{2020}]{zhao2020}
Zhao D., Du M., Ho L.~C., Debattista V.~P., Shi J., 2020, ApJ, 904, 170

\bibitem[\protect\citeauthoryear{Zhou et al.}{2020}]{zhou2020}
Zhou Z.-B., Zhu W., Wang Y., Feng L.-L., 2020, ApJ, 895, 92.



\end{thebibliography}
%

\begin{appendix} 
\section{Kinematic decomposition and bar identification}
\label{app:decomp}
In order to identify more precise galaxy components, we adopt the kinematic decomposition algorithm \textsc{mordor}\footnote{ \url{https://github.com/thanatom/mordor}} presented in  \cite{zana2022} to determine more specific galaxy components.  The decomposition is predicated on the circularity ($\epsilon$) and binding energy ($E$) phase space, where a minimum in $E$ is identified for each galaxy, $E_{\rm cut}$. All four components listed below  were identified  for the simulations employed in the paper:
\begin{itemize}
    \item \textbf{Classical Bulge:} is defined as the set of stellar particles that exhibit the highest degree of binding, with a value that is contingent upon the $E_{\rm cut}$ of each galaxy. These particles also exhibit counterrotation, which is characterised by a negative value of $\epsilon$ ($<0$).  Next, Monte Carlo sampling is employed to select a distribution that is equivalent to the component of the distribution that exhibits positive circularity and allocate it to the Bulge.
    \item \textbf{Thin/Cold disc:} is defined as the set of stellar particles that have the highest degree of binding and are not allocated to the Bulge, but have positive values of $\epsilon$ ($\epsilon>0.7$).
    \item \textbf{Pseudobulge:} is defined as the remaining stellar particles that exhibit a high degree of binding without being assigned to the Bulge or the thin disc.
    \item \textbf{Thick/Warm disc:} is defined as the set of stellar particles that have positive $\epsilon$ values ($\epsilon>0.7$) and are not assigned to the bulge/pseudobulge, despite exhibiting a reduced degree of binding.
\end{itemize}

\begin{figure*}

\includegraphics[width=2\columnwidth]{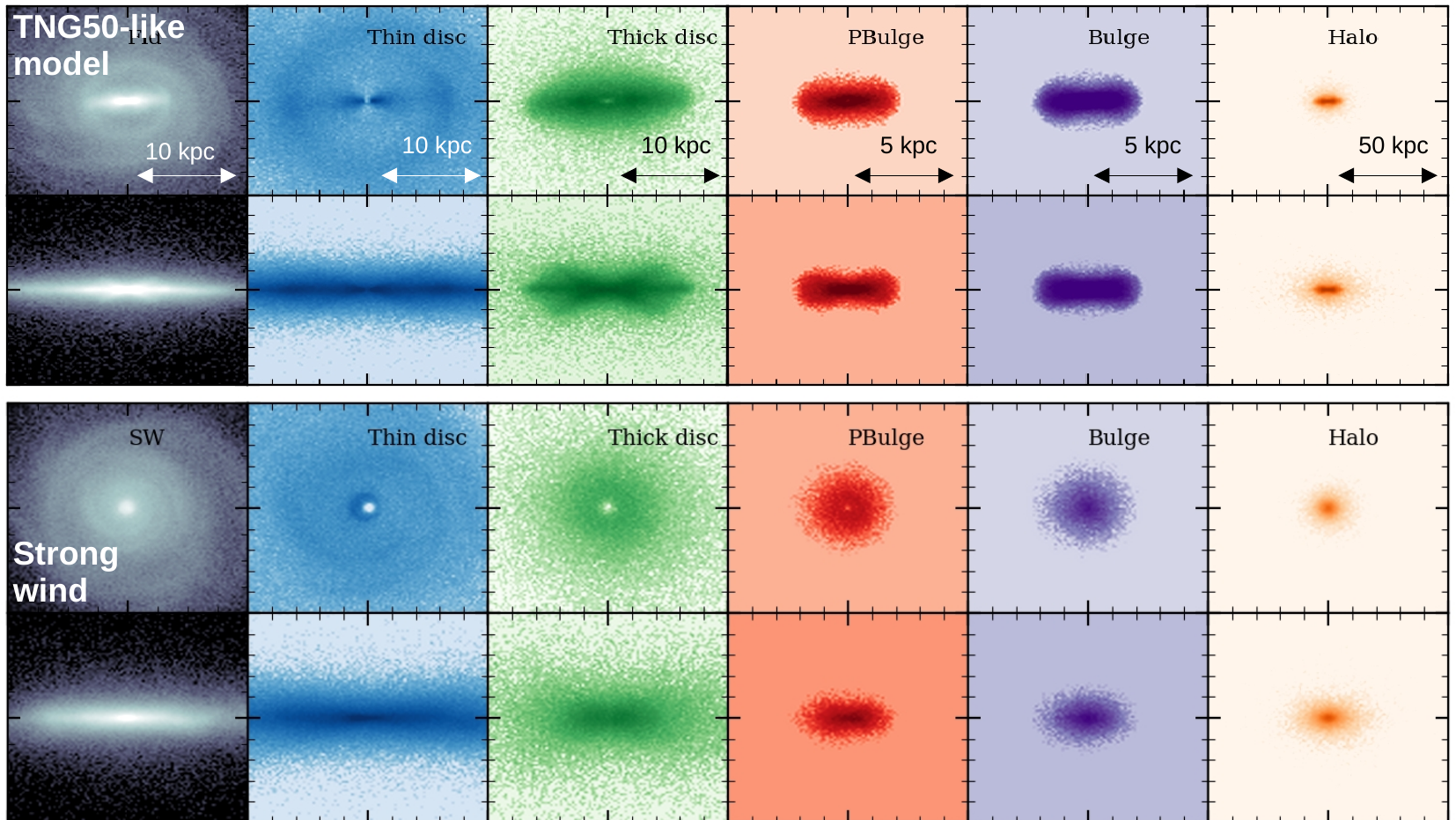}
\caption{Stellar density maps of the galaxy components identified by \textsc{mordor}.From left to right Cols.: total galaxy, thin disc, thick disc,  Pseudobulge, bulge, and halo. Top panels correspond to a galaxy in the TNG50-like model. Bottom panel to the same galaxy in the strong wind model.}
\label{fig:mordor}
\end{figure*}

The stellar density maps components of a galaxy are illustrated in Fig. \ref{fig:mordor} for the TNG50-like model, in which the galaxy forms a bar, and for the strong wind model, in which the galaxy does not form a bar. \textsc{mordor} is employed to generate these maps. The distinctions between all of the components are readily apparent.  For this example, the thick disc seems to trace the bar, while the pseudobulge appears to be indistinguishable from the bulge. This also seems to be the case at the time of bar formation.

\subsection{Identification of bars}
\label{app:bars}
The face-on stellar surface density is Fourier decomposed to identify non-axisymmetric structures (e.g. \citealt{athanassoula2002,zana2018a,rosasguevara2020}).
We concentrate on $\Ato(R)$, which is defined by the ratio between the second and zero terms of the Fourier expansion and its phase $\Phi(R)$ (refer to equations 1 and 2 in \citetalias{rosasguevara2022}). A bar structure is characterised by $\Ato(R)$ and $\Phi(R)$, with the bar strength defined as $\Amax$, the peak of $\Ato(R)$. Inside the bar, the phase should be constant. To establish a constant phase, we calculate the standard deviation ($\sigma$) of $\Phi(R)$ every time a new cylinder shell is added and set $\sigma\leq 0.1$.  The bar extent ($\rbar$) is the greatest radius where $\sigma\leq 0.1$, and the $\Ato$ profile first dips below $0.15$ or the minimum value of $\Ato(R)$.
Large values of $\Ato(R)$ may be attributable to transient events like mergers and interactions; thus, we assume the bar represents a long-lasting characteristic if
\begin{enumerate}
\item The maximum of $\Ato$, $\Amax$, is greater than $0.2$,
\item $\rbar>r_{\rm min}$ where $r_{\rm min}=1.38\times\epsilon_{*,z}$ is a  minimum radius imposed and $\epsilon_{*,z}$ corresponds to the proper softening length for stellar particles.
\item The estimated age of the bar is larger than the time between the analysed output and 2 previous simulation outputs.
\end{enumerate}

Although this filter may omit newer bars at lower redshifts, it aligns with the bar selection criteria employed by \citetalias{rosasguevara2020,rosasguevara2022} when bar structure was found in previous outputs.

To identify a bar structure, the formation time, $\tbar$, must also be determined. We follow the growth of $\Amax$ up to the moment when $\Amax \leq 0.2$ and $\rbar(\tbar)\leq r_{\rm min}$ for several snapshots.  Additionally, the difference between $\Amax$ at a snapshot and the two prior snapshots must not exceed $0.45$ during this time frame. This confirms that the bars we find are stable.

\subsection{Method of three-component decomposition of surface face-on density profile}
In this section, we outline the process of decomposing the surface brightness profiles of the galaxies to determine the length scale of the disc, the effective radius of the bulge, and the properties of the bar when it is identified through Fourier decomposition.
The surface density profiles are calculated in a face-on perspective within concentric annuli that are $0.16$ dex in width and centred on the stellar component minimal potential.

We employ the particle swarm optimisation (PSO) code PSOBacco \footnote{ \url{https://github.com/hantke/pso_bacco}} \citep{arico2021} to determine the minimum$\chi^{2}$  of a fit of the sum of an exponential profile and two Sersic profiles (one of which corresponds to the bar). This value is expressed in terms of the stellar mass as

\begin{equation}
\begin{aligned}
\Sigma(R) &=&\Sigma_{\rm d,0}\, {\rm exp} \bigg (- R/\hdisc \bigg ) + \Sigma_{\rm b,e} \,{\rm exp}\bigg ( -b_{n} \bigg [(r/r_{\rm b,eff})^{1/n} -1  \bigg ] \bigg ) \\
          & &
+\Sigma_{\rm bar,e} \,{\rm exp}\bigg ( -b_{n} \bigg [(r/r_{\rm bar,eff})^{1/n_{\rm bar}} -1  \bigg ] \bigg )
\end{aligned}
\label{eq:sigma}
\end{equation}
where $\Sigma_{\rm d,0}$ is the central surface density of the disc component, $\hdisc$ is the disc scale length,  $r_{b,\rm eff}$  ($r_{\rm bar}$) the effective radius that encloses half of the stellar mass of the  one predicted from the Sersic profile,  $\Sigma_{\rm b,e} $ ($\Sigma_{\rm bar,e} $) corresponds to the surface density at $r_{b,\rm eff}$, and $n$ is the Sersic index (of the bulge/bar). The value of $b_{n}$ is contingent upon the complete gamma function, as  $\Gamma(2n)= 2\gamma(2n,b_{n})$.

We apply the method of fitting three components in the region at radii smaller than $r_{\rm fit} = \rm max (\rm log_{10}(\Sigma(r)/\Sigma_{\rm max})\geq -2.6,10\,\kpc)$ when a bar is detected by Fourier decomposition. If not, we fit two components. This condition guarantees to suit both high-redshift discs, which are more compact (smaller disc scale lengths for a given stellar mass), and their analogues at low redshifts, whose discs are less compact (larger disc scale lengths for a given stellar mass).

\begin{figure*}
\begin{tabular}{cc}
\includegraphics[width=0.75\columnwidth]{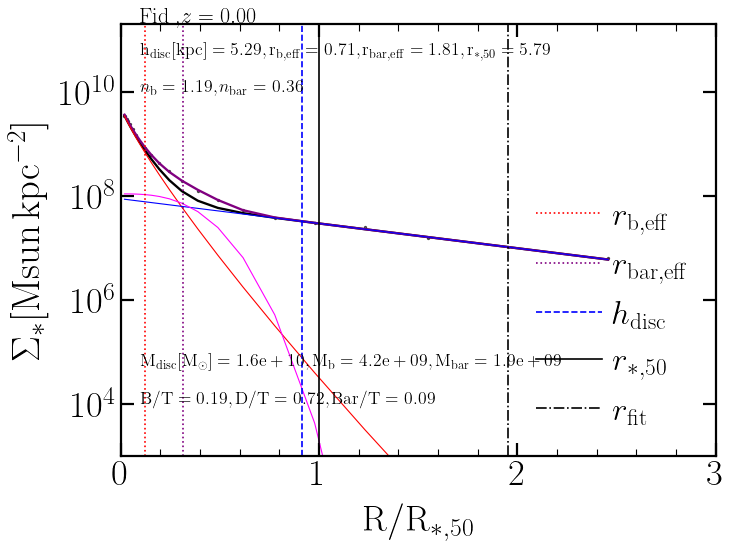} &
\includegraphics[width=0.75\columnwidth]{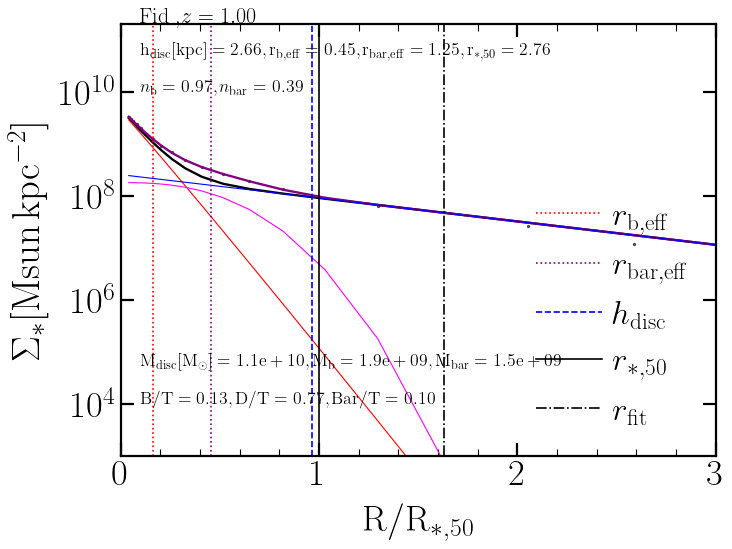}\\
\includegraphics[width=0.75\columnwidth]{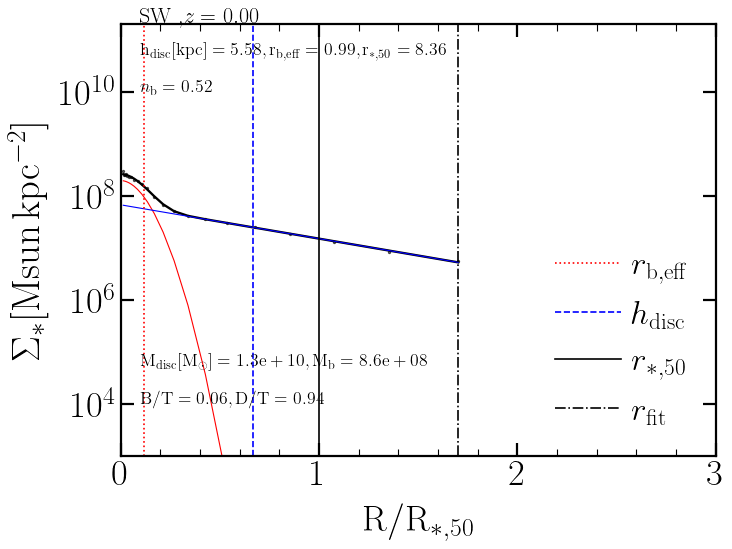} &
\includegraphics[width=0.75\columnwidth]{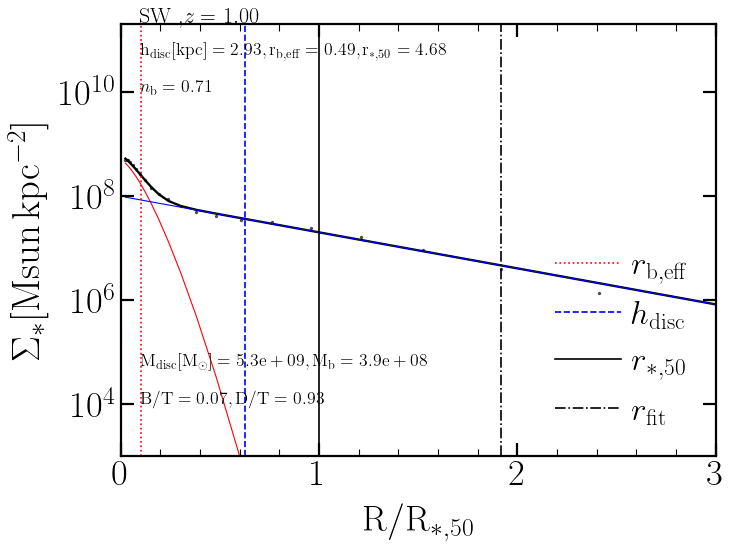}    \\

\end{tabular}
\caption{Face-on stellar surface density profiles in terms of  $R/R_{*,50}$ where $R_{*,50}$ is the stellar half-mass radius, for a galaxy with a bar in the TNG50-like model (top) and unbarred galaxy in the strong wind model (bottom) at $z=0,1$. The profiles (black lines) were
obtained by fitting simultaneously a Sersic profile for a bulge (red lines) and a Sersic for a bar (pink lines), and an exponential profile (blue lines). Purple lines represent the surface density profile from the predicted Sersic profile and exponential profile. Horizontal dashed lines correspond to the radius at which the fit is done. Dotted horizontal lines correspond to the bar length (pink), the effective radius of the bulge (red), and the scale length of the disc(blue).}
\label{fig:fitexamples}
\end{figure*}

The surface density profiles and the fits for a galaxy in the TNG50-like model, where a bar is formed, and the same galaxy in the strong wind model, where a bar is not formed, are illustrated in Fig. \ref{fig:fitexamples}. A black line denotes the total fitted profile. The disc, bar, and bulge components are represented by blue, pink, and red lines, respectively. As a point of reference, the scale-length of a disc, $\hdisc$, the effective radius of the bulge, $r_{\rm b,eff}$, and the length of the bar in the case of barred galaxies are represented by coloured vertical lines. The vertical lines that are dashed represent the radius at which the fit is performed.

\end{appendix}

\end{document}